\documentclass{emulateapj}
\usepackage[breaklinks,colorlinks,urlcolor=blue,citecolor=blue,linkcolor=blue]{hyperref}
\usepackage{mathrsfs}
\usepackage{amsmath}
\usepackage{graphicx}

\shortauthors{Metzger et al.}

\shorttitle{Extragalactic Transients in the Era of Wide-Field Radio
  Surveys. I}

\newcommand\degd{\ifmmode^{\circ}\!\!\!.\,\else$^{\circ}\!\!\!.\,$\fi}

\newcommand{\swsixteen}{\protect\objectname[Swift J164449.3+573451]{Sw\,J1644$+$57}}

\begin{document}

\title{Extragalactic Transients in the Era of Wide-Field Radio
  Surveys. I. Detection Rates and Light Curve Characteristics}

\author{
Brian D.~Metzger\altaffilmark{1},
P.~K.~G.~Williams\altaffilmark{2},
\& Edo Berger\altaffilmark{2}
}

\altaffiltext{1}{Columbia Astrophysics Laboratory, Pupin Hall, New
  York, NY, 10027, USA; bmetzger@phys.columbia.edu}

\altaffiltext{2}{Harvard-Smithsonian Center for Astrophysics, 60
  Garden Street, Cambridge, MA 02138; pwilliams@cfa.harvard.edu,
  eberger@cfa.harvard.edu}

\begin{abstract}
  The impending era of wide-field radio surveys has the potential to
  revolutionize our understanding of astrophysical transients. Here we
  evaluate the prospects of a wide range of planned and hypothetical
  radio surveys using the properties and volumetric rates of known and
  hypothetical classes of extragalactic synchrotron radio transients
  (e.g., on- and off-axis gamma-ray bursts, supernovae, tidal
  disruption events, compact object mergers).  Utilizing these sources
  and physically motivated considerations we assess the allowed
  phase-space of radio luminosity and peak timescale for extragalactic
  transients.  We also include for the first time effects such as
  redshift evolution of the rates, K-corrections, and non-Euclidean
  luminosity distance, which affect the detection rates of the most
  sensitive surveys.  The number of detected events is calculated by
  means of a Monte Carlo method, using the various survey properties
  (depth, cadence, area) and realistic detection criteria that include
  a cut on the minimum variability of the transients during the survey
  and an assessment of host galaxy contamination. For the detected
  events we also quantify how well each light curve is characterized
  (e.g., the fraction of sources with a measured rise time, decline
  time, and peak brightness). We find that near-term GHz frequency
  surveys (ASKAP/VAST, Very Large Array Sky Survey) will detect few
  events: $\lesssim 30-50$ on- and off-axis long GRBs and off-axis
  tidal disruption events, and $\sim 10-20$ neutron star binary
  mergers if $\sim 1\%$ of the mergers result in a stable millisecond
  magnetar. Low-frequency surveys (e.g., LOFAR) are unlikely to detect
  any transients, while a hypothetical large-scale mm survey may
  detect $\sim 40$ on-axis long GRBs.  On the other hand, we find that
  SKA surveys at $\sim 0.1-1$ GHz have the potential to uncover
  thousands of transients, mainly on- and off-axis long GRBs, on-axis
  short GRBs, off-axis TDEs, and neutron star binary mergers with
  magnetar remnants.
 \end{abstract}


\section{Introduction}
\label{sec:intro}

Radio astronomy appears poised for a revolution in the study of
transient phenomena thanks to the advent of wide-field
interferometers. At meter wavelengths (tens to hundreds of MHz) it is
now feasible to image the full primary beams of dipole antennae,
resulting in new advanced arrays such as the LOw Frequency ARray
(LOFAR; \citealt{VanHaarlem+13}), the Murchison Widefield Array (MWA;
\citealt{Lonsdale:2009aa}) and the Long Wavelength Array (LWA;
\citealt{Ellingson:2009aa}). At centimeter wavelengths (GHz
frequencies) a new generation of wide-field facilities is also being
developed, such as MeerKAT \citep{Booth:2009aa} and the Australian
Square Kilometer Array Pathfinder (ASKAP; \citealt{Johnston:2008aa}),
and existing facilities are undergoing upgrades, including the Apertif
project at the Westerbork Synthesis Radio Telescope (WSRT;
\citealt{Oosterloo:2009aa}) and the expansion of the Karl G.~Jansky
Very Large Array (VLA; \citealt{Perley+11}). The VLA upgrade makes it
possible to reach greater image depth in a shorter time, raising the
possibility of conducting a new survey combining the sensitivity of
the FIRST survey \citep{bwh95} with the large sky footprint of the
NVSS survey \citep{ccg+98}. These telescopes and new technologies are
ultimately paving the way for the Square Kilometer Array (SKA;
\citealt{cr04}), which will further transform radio astronomy with its
unprecedented scale and sensitivity.

Despite the overall maturity of radio astronomy, surprisingly little
is known about what astrophysical sources might dominate the transient
radio sky.  At long wavelengths, the brightest sources are expected to
result from coherent emission processes. Examples include `giant
pulses' from Galactic pulsars (e.g.~\citealt{Jessner+05}) and
cyclotron maser emission from brown dwarfs and possibly exoplanets
\citep{Berger:2002aa,Hallinan:2008aa}; brighter extra-galactic analogs
of such events are speculated to accompany rare violent events, such
as giant magnetar flares (\citealt{Lyubarsky:2014aa}) and mergers of
binary neutron stars \citep{Hansen:2001aa}, and may be connected to
the fast radio burst (FRB) phenomenon (e.g., \citealt{Thornton:2013aa,
  Spitler:2014aa}). At centimeter wavelengths, and on longer
timescales, synchrotron sources are expected to dominate the transient
sky. These are generally powered by the shock interaction between fast
ejecta from an energetic explosion and dense ambient gas, as in the
case of supernovae (SNe) and gamma-ray bursts (GRBs).  Slow-timescale
Galactic radio transients are expected to be dominated by a different
population, such as flares from M dwarfs and X-ray binaries, with
rates that are believed to be comparable to those of extragalactic
events \citep{Williams+13}.

In this work, we consider and study extragalactic synchrotron
transients. These are the best understood events based on radio
follow-up of discoveries at other wavelengths, they are of proven
astrophysical interest, and they are amenable to detection by imaging
surveys, especially those operating at centimeter wavelengths. Known
examples include core-collapse SNe \citep{Kulkarni+98, Weiler+02,
  Berger+03, Soderberg+10}, GRB afterglows \citep{fkn+97,
  Chandra&Frail12}, and relativistic jets from tidal disruption events
\citep{Giannios&Metzger11, Zauderer+11, Berger+12, Zauderer+13}.

Searches for these and related types of radio transients help to
motivate the science cases for all of the telescopes and surveys
mentioned above.  With these next generation projects coming online in
the near future, it is essential to consider how many and which kinds
of transient events will be detected, and how survey strategies may be
optimized to answer particular science questions.  Most untargeted
surveys to date have detected few if any events (e.g., \citealt{cif03,
  cba+10errated, bfs+11}), and despite careful vetting, a substantial
number of detected candidates have turned out to be instrumental
artifacts \citep{Gal-Yam+06, Ofek+10, cbk+11, Frail+12}.  Thus, most
of what is known empirically about the transient radio sky consists of
upper limits.  Complicating matters is the fact that these upper
limits are often difficult to characterize and interpret, because the
expected yield of a given survey results from a complex interplay
between the properties of the transient events in question and the
survey strategy. While this interplay is often considered in the
derivation of upper limits and event rates \citep{Bower+07, cbk+11,
  mfo+13, Williams+13}, it is challenging to harmonize the results
from different studies for inter-comparison, or to extrapolate them to
new surveys.

While the current results from untargeted surveys are of limited use
for predicting the yields of the next-generation projects, all is not
lost since some information is available from radio follow-up
observations and from generic theoretical considerations. The
volumetric rates of some of the expected events have been explored in
detail (e.g., \citealt{Wang&Merritt04, Guetta&DellaValle07,
  Wanderman&Piran10, Kim+13}), and targeted radio follow-up of
transients triggered from other wavelengths has yielded significant
insight into the radio properties of some of the event classes
mentioned above. In other cases, detailed numerical models allow
predictions of radio emission that are well-motivated by both
observations and theory (e.g., \citealt{Nakar&Piran11,
VanEerten&MacFadyen11}). 

Despite this available knowledge, the predicted rates of radio
transient occurrence and detection --- taking into account the
expected luminosities and durations --- have generally not received
detailed theoretical attention. \citet{Frail+12} estimated event rates
ignoring cosmological effects such as non-Euclidean luminosity
distances, K-corrections, and time dilation, instead assuming the
standard isotropic Euclidean scaling $N_{\rm all-sky}(>F_{\nu})\propto
F_\nu^{-3/2}$, where $N_{\rm all-sky}$ is the instantaneous number of
events on the sky brighter than a given flux density, $F_\nu$. Other
works considered specific event classes with more detailed analyses
\citep{Ghirlanda+14, Feng+14arxiv} but do not treat the surveys
themselves in detail.  As we show in this work, the interplay between
the transient luminosities and timescales, the survey cadence, and the
total survey duration is critical to determining the yield of a given
survey, and the ability to separate transient from steady sources.

Here we offer the first detailed and comprehensive assessment of the
discovery prospects for extragalactic synchrotron radio transients in
the upcoming wide-field survey era.  We do so by considering multiple
effects: the specific observing strategies of several proposed and
hypothetical surveys, the light curves for a wide range of known and
theoretical transients spanning a broad range of energies,
collimation, and ambient densities (and hence timescales and
luminosities), a judicious choice of detectability metrics, and
realistic volumetric rates and their cosmological evolution. Our goal
is to provide well-motivated estimates for what various surveys will
actually discover using the best information currently
available. Future radio surveys also have the potential to discover
new classes of transients beyond the ones we consider here.  However,
any transient source powered by synchrotron radiation from a rapidly
expanding blastwave obeys fundamental physical constraints, such as
the relation between peak luminosity and timescale, and a
characteristic evolution from high to low frequencies.  Thus, our
investigation of a broad range of the synchrotron transient parameter
space sets meaningful bounds on the possible detection rates even of
event classes that we do not explicitly consider.

Our primary outputs in this paper are the numbers and basic light
curve characteristics of detected transients as a function of event
class and survey.  In addition to this information it is also critical
to consider what can (and cannot) be learned and extracted from each
event after detection, such as the energy scale, collimation, ambient
density, and nature of the transient, These will depend in part on the
ability to robustly measure the peak luminosity, characteristic
timescale, and redshift from a host galaxy. In a follow-up paper
(``Paper II'') we will investigate this topic, combining the radio
light curves produced by our simulations from this paper with
synchrotron models and realistic follow-up strategies.  Our analysis
will be conducted with an eye towards optimizing future transient
surveys to yield not just detections but also astrophysical insight.

The paper is organized as follows. We begin by considering the general
characteristics of radio emission from extragalactic synchrotron
transients (\S\ref{sec:general}), and establishing the particular
properties of the events that we consider in this paper, including
luminosities, timescales, and volumetric rates (\S\ref{sec:models}).
In \S\ref{sec:surveys} we describe the characteristics of the radio
surveys that we simulate.  Our Monte Carlo method and detection
criteria for the simulations are described in \S\ref{sec:monte}, and
their results are summarized in \S\ref{sec:results}.  We discuss the
implications of our results and present our conclusions in
\S\ref{sec:discussion}.

\begin{deluxetable*}{lccccccccc}
\tabletypesize{\scriptsize}
\tablecolumns{6}
\tabcolsep0.05in\footnotesize
\tablewidth{0pc}
\tablecaption{Extragalactic Transient Classes
\label{tab:classes}}
\tablehead {
\colhead{Transient}                 &
\colhead{$\mathcal{R}(z=0)$}  &
\colhead{$E_K$}              &
\colhead{$n$}                 &
\colhead{$\beta_i$}        &
\colhead{$\tau_{\rm 0.15}\,^{(a)}$}      &
\colhead{$\tau_{\rm 1.3}\,^{(b)}$}       &
\colhead{$\tau_{\rm 3}\,^{(c)}$}          &
\colhead{$\tau_{\rm 150}\,^{(d)}$}       &
\colhead{Ref.}                                 \\
\colhead{}           &
\colhead{(Gpc$^{-3}$ yr$^{-1}$)}  &
\colhead{(erg)}            &
\colhead{(cm$^{-3}$)}  &
\colhead{}               &
\colhead{(d)}           &
\colhead{(d)}           &
\colhead{(d)}           &
\colhead{(d)}          &
\colhead{}
}
\startdata
LGRB, $\theta_{\rm obs} = 0.2$ & 0.3$^{\dagger}$ & $10^{51}$ & 1 & 1 & 320 & 62 & 31 & 4 & 1 \\
LGRB, $\theta_{\rm obs} = 0.4$ & 1$^{\dagger}$ & $10^{51}$ & 1 & 1 & 450 & 90 & 55 & 12 &  1 \\
LGRB, $\theta_{\rm obs} = 0.8$ & 4$^{\dagger}$ & $10^{51}$ & 1 & 1 & 1900 & 230 & 150 & 150 & 1   \\
LGRB, $\theta_{\rm obs} = 1.57$ & 12$^{\dagger}$ & $10^{51}$ & 1 & 1 & 1300 & 620 & 550 & 590 & 1  \\
Low Luminosity LGRB (``LLGRB'') & 500$^{\dagger}$ & $10^{49}$ & 1 & 0.8 & 200 & 43 & 90 & 110 & 9  \\
SGRB, $\theta_{\rm obs} = 0.2$ & 5$^{\dagger}$ & $10^{50}$ & $10^{-3}$ & 1 & 220 & 110 & 90 & 110 & 6\\
SGRB, $\theta_{\rm obs} = 0.4$ & 15$^{\dagger}$ & $10^{50}$ & $10^{-3}$ & 1 &  360 & 180 & 160 & 180 & 6 \\
SGRB, $\theta_{\rm obs} = 0.8$ & 60$^{\dagger}$ & $10^{50}$ & $10^{-3}$ & 1 &  730 & 480 & 410 & 480 & 6 \\
SGRB, $\theta_{\rm obs} = 1.57$ & 185$^{\dagger}$ & $10^{50}$ & $10^{-3}$ & 1 & 1900 & 2200 & 2000& 650 & 6 \\
On-Axis TDE (``\swsixteen") & 0.01$^{\ddagger}$ & $10^{52}$ & 0.1 & 1 & 3700 & 920 & 1040 & 180 & 2--4 \\
Off-Axis TDE, spherical & 1$^{\ddagger}$ & $10^{52}$ & 0.1 & 0.8 & 3700 & 900& 900 & 900 & 5  \\
NSM: prompt BH & $500$$^{\dagger}$ & $3\times 10^{50}$ & 0.1 & 0.2 & 4000 & 4000 & 4000 & 4000 & 7\\
NSM: stable remnant (``NSM-magnetar'') & $5$$^{\dagger}$  & $3\times 10^{52}$ & 0.1 & 1 & 2800 & 1300 & 1300 & 1300 & 8 \\
Type Ib/c SNe (``RSN") & 5000$^{\dagger}$ & $10^{47}$ & 1 & 0.2 & 870 & 120 & 55 & 1.1 & 10
\enddata
\tablecomments{$^{\dagger}$Scaled with redshift according to star
  formation rate (\citealt{Cucciati+12}). $^{\ddagger}$ Scaled with
  redshift according to volumetric density of supermassive black holes
  (\citealt{Sijacki+14}). $^{(a)}$ Light curve duration at observer
  frequency $\nu = 0.15$ GHz. $^{(b)}$ Light curve duration at observer
  frequency $\nu = 1.3$ GHz. $^{(c)}$ Light curve duration at observer
  frequency $\nu = 3$ GHz. $^{(d)}$ Light curve duration at observer
  frequency $\nu = 150$ GHz.
  References: (1) \citealt{VanEerten+10};
  (2) \citealt{Zauderer+11}; (3) \citealt{Berger+12}; (4)
  \citealt{Zauderer+13}; (5) \citealt{Giannios&Metzger11}; (6)
  \citealt{VanEerten&MacFadyen11}; (7) \citealt{Nakar&Piran11}; (8)
  \citealt{Metzger&Bower14}; (9) \citealt{BarniolDuran+14}; (10)
  \citealt{Soderberg+08} }
\end{deluxetable*}

\section{Extragalactic Transients}
\label{sec:classes}

In this work we focus on extragalactic synchrotron-emitting
transients, powered by shock interaction between fast ejecta and the
ambient medium. The production of synchrotron emission in such
transients is generic, and has been observed from various sources
(e.g., GRBs, SNe, TDEs).  Here we consider a wide range of transients
in terms of their energy scale and ambient density that include both
relativistic and non-relativistic explosions, collimated and spherical
outflows, with on-axis and off-axis orientations. As a result, these
transients span a broad range of the luminosity-timescale phase-space
that is commonly used to map radio transients; see
Figure~\ref{fig:Tc}.  We note that some of these transients have been
extensively observed in the past using radio follow-up of discoveries
at other wavelengths (e.g., on-axis GRBs, SNe), some include only a
few known examples (e.g., TDEs), some are robustly predicted but have
not been directly observed to date (e.g., binary neutron star mergers,
off-axis GRBs), and some are purely hypothetical (e.g., neutron star
mergers giving rise to a stable millisecond magnetar). While other
types of extragalactic synchrotron radio transients could be
hypothesized, the sources considered here span a sufficiently broad
range of properties and volumetric rates to encompass future
predictions.

\subsection{General Considerations of Timescale and Luminosity for
  Extragalactic Radio Transients}
\label{sec:general}

For the purpose of placing general constraints on the light curves of
extragalactic synchrotron radio transients we begin by considering the
ejection of material with a kinetic energy of $E_K= 10^{51}\,E_{K,51}$
erg and an initial velocity of $v_i=\beta_i c $ (corresponding to an
initial Lorentz factor of $\Gamma_i= (1-\beta_i^{2})^{-1/2}$), into an
ambient medium of constant density $n=1\,n_{0}$ cm$^{-3}$. The ejecta
transfer their energy to the ambient medium at the characteristic
radius ($R_{\rm dec}$) at which point they have swept up a mass
comparable to $\sim 1/\Gamma_i$ of their rest mass:
\begin{eqnarray}
  R_{\rm dec} &\approx& \left(\frac{3E_K}{4\pi n m_p c^{2} \Gamma_i^{2}
      \beta_i^2} \right)^{1/3} \nonumber \\
&\approx& 6\times 10^{17}\,{\rm cm}\,\,
  E_{K,51}^{1/3}n_{0}^{-1/3}\Gamma_i^{-2/3}\beta_i^{-2/3}.
\label{eq:rdec}
\end{eqnarray}
This occurs at the deceleration timescale:
\begin{equation}
  t_{\rm dec} \approx R_{\rm dec}/2c\beta_i\Gamma_i^{2} \approx
  115\,{\rm d}\,\,E_{K,51}^{1/3}n_{0}^{-1/3}\beta_i^{-5/3}
  \Gamma_i^{-8/3}.
\label{eq:tdec}
\end{equation} 
Prior to $t_{\rm dec}$ the radio brightness will rise, so this
timescale defines a minimum peak time for radio transients (although
the peak time could be longer; see below).

If the observing frequency ($\nu_{\rm obs} = 1\,\nu_{\rm GHz}$ GHz) is
located above both the synchrotron peak frequency ($\nu_m$) and the
self-absorption frequency ($\nu_a$), then the peak brightness is
achieved at $t_{\rm dec}$, and is given by \citep{Nakar&Piran11}:
\begin{equation}
  F_{\nu,\rm dec} \approx 0.05\,{\rm mJy}\,\, E_{K,51} n_{0}^{0.83}
  \epsilon_{e,-1}^{1.3}\epsilon_{B,-2}^{0.83}\beta_i^{2.3}
  D_{L,28}^{-2}\nu_{\rm GHz}^{-0.65}.
\label{eq:Fp}
\end{equation} 
Here $D_L = 10^{28}\,D_{L,28}$ cm is the luminosity distance.  We also
make the standard assumption that electrons are accelerated at the
shock front into a power-law energy distribution $N(E)\propto E^{-p}$
with $p=2.3$ above a minimum Lorentz factor of $\gamma_m$, and that
$\epsilon_B = 0.01\,\epsilon_{B,-2}$ and $\epsilon_e= 0.1\,
\epsilon_{e,-1}$ are the fractions of post-shock energy in the
magnetic field and relativistic electrons, respectively.

Some of the transients we consider in this work (e.g., SNe, binary
neutron star mergers) produce only non- or mildly-relativistic ejecta
and are essentially spherical with isotropic emission at all times.
Other transients (e.g. GRBs, jetted TDEs) produce ultra-relativistic
ejects in tightly collimated jets\footnotemark\footnotetext{GRB jets
  have $\Gamma_i\gtrsim 100$ and $\theta_j\sim 0.1$ (e.g.,
  \citealt{Frail+01}), while TDE jets have $\Gamma_i\sim 10$ and
  $\theta_j \sim 0.1$ (e.g., \citealt{Metzger+12,Berger+12}).}  with
opening angles of $\theta_ j\ll 1$. However, when viewed in an initial
off-axis direction, the emission from these sources can also be
approximated as being spherically symmetric once the shocked matter
decelerates to sub-relativistic velocities and spreads laterally into
the observer's line of sight (e.g., \citealt{Zhang&MacFadyen09};
\citealt{Wygoda+11}).  At this point the radio emission is no longer
strongly beamed, and $t_{\rm dec}$ and $F_{\nu,\rm dec}$ can be
approximated using $\beta_i\approx 1$ \citep{Nakar&Piran11}.

From Equations~\ref{eq:tdec} and \ref{eq:Fp} it is clear that the
luminosities of synchrotron transients scale linearly in proportion to
their kinetic energy, but that more energetic events also evolve on a
slower characteristic timescale, $t_{\rm dec}\propto E_K^{1/3}$. The
timescale can be further increased by cosmological time dilation,
$t_{\rm dec} \propto (1+z)$, if such events are energetic enough to be
detectable at substantial redshifts.

Explosions in high density environments are also more luminous
(Equation~\ref{eq:Fp}) but exhibit shorter timescales ($t_{\rm dec}
\propto n^{-1/3}$). High densities could thus in principle lead to
very luminous short-duration transients. However, as the density
increases synchrotron self-absorption becomes important and eventually
suppresses the brightness and leads to a longer duration because the
relevant timescale to reach peak brightness is no longer $t_{\rm dec}$
but instead the timescale at which $\nu_a=\nu_{\rm obs}$, i.e., when
the optical depth is of order unity. The self-absorption frequency at
$t_{\rm dec}$ is given by \citep{Nakar&Piran11}:
\begin{equation}
  \nu_a (t_{\rm dec}) \approx 0.8\,{\rm GHz}\,\,E_{K,51}^{0.11}
  n_{0}^{0.49}\epsilon_{B,-2}^{0.34}\epsilon_{e,-1}^{0.41}
  \beta_i^{1.3}.
\end{equation}
Thus, self-absorption may play a significant role in suppressing the
brightness and extending the timescale of transients, especially at
sub-GHz frequencies. On the other hand, the steep self-absorbed
spectrum ($F_\nu\propto \nu^{2}$ or $\nu^{5/2}$) can lead to a
positive K-correction for sources that are energetic enough to be
detectable at cosmological distances (e.g., on-axis GRB afterglows;
\citealt{Frail+06}).

Finally, we note that $t_{\rm dec}$ can be very short for relativistic
sources ($t_{\rm dec}\propto \Gamma_i^{-8/3}$) so in principle such
sources can exhibit short timescales. However, for relativistic
sources the relevant peak timescale {\it at radio frequencies} is not
determined by $t_{\rm dec}$ because on this timescale the synchrotron
peak frequency generically obeys $\nu_m\gtrsim \nu_{\rm obs}$. The
value of $\nu_m$ is determined by the Lorentz factor of the minimum
energy electrons, $\gamma_m \approx 40\epsilon_{e,-1} \Gamma$
\citep{spn98}, such that for large values of $\Gamma$ it is located
above GHz frequencies until the outflow decelerates significantly.
Thus, the radio light curve at $\nu_{\rm obs}$ will reach peak
brightness only when $\nu_m=\nu_{\rm obs}$ (as long as $\nu_a\lesssim
\nu_m$; otherwise the peak will occur when $\nu_a=\nu_{\rm obs}$). The
value of $\nu_m$ for relativistic sources in a constant density medium
(for $\nu_m\gtrsim \nu_a$) is given by \citep{gs02}:
\begin{equation}
  \nu_m \approx 1.0\,{\rm GHz}\,\, E_{\rm K,iso,53}^{1/2}
  \epsilon_{e,-1}^{2} \epsilon_{B,-2}^{1/2} t_2^{-3/2},
\end{equation}
where $E_{\rm K,iso}$ is the isotropic-equivalent energy, which takes
into account the collimation of relativistic sources, the scaling to
$10^{53}$ erg is used since a typical beaming correction of $\sim
0.01$ will give a fiducial $E_K\approx 10^{51}$ erg, and
$t=100\,t_2$~d is the time since explosion.  Thus, relativistic
transients with $\Gamma_i\gg 1$ have a timescale for peak emission at
radio frequencies that is much longer than $t_{\rm dec}$.

\begin{figure*}
\includegraphics[width=\textwidth] {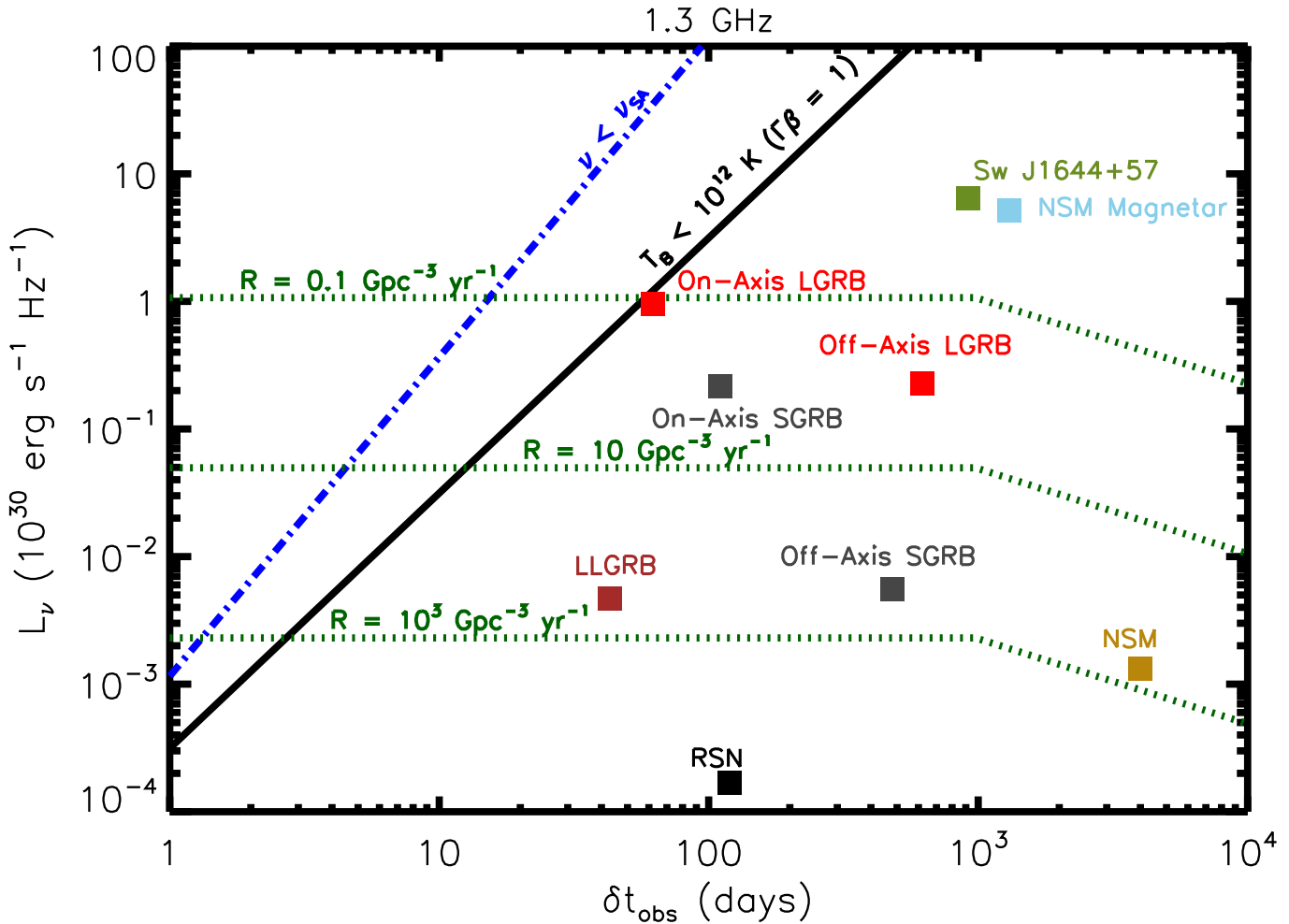}
\caption{Phase-space of peak luminosity and timescale for transient
  synchrotron sources at a fiducial frequency of $1.3$ GHz.  Sources
  in the region of phase-space above the solid black line will violate
  the brightness temperature constraint $T_B\lesssim 10^{12}$ K
  (Equation~\ref{eq:FnuTC}).  An independent constraint on the allowed
  phase space is given by the condition that $\nu_m,\nu_a\lesssim
  \nu_{\rm obs}$ (Section~\ref{sec:general}; dot-dashed blue
  line). The green dotted lines indicate the minimum volumetric rate
  of transients required for at least possible one detection across
  the entire sky for a three-year survey with a limiting sensitivity
  of $1$ mJy (\S\ref{sec:general}).  Transients with lower peak
  luminosities (and hence potentially shorter peak timescales) can
  only be detected within smaller volumes and therefore require higher
  intrinsic rates.  These minimum rates can be compared to actual
  transient rates as summarized in Table~\ref{tab:classes}.  Finally,
  colored squares denote the peak luminosity and timescale of the
  various transients considered in this paper: on- and off-axis LGRBs
  ({\it red}), on- and off-axis short SGRBs ({\it gray}), neutron star
  binary mergers with prompt black hole formation (NSM: {\it tan}),
  neutron star binary mergers with a magnetar remnant (NSM magnetar:
  {\it blue}), on-axis jetted TDEs (\swsixteen: {\it light green}),
  low luminosity GRB (LLGRB: {\it brown}), and Type Ib/c SNe
  (RSN: {\it black}). We note that all of these sources indeed obey
  the basic luminosity-timescale limits discussed in
  \S\ref{sec:general}.}
\label{fig:Tc}
\end{figure*}

To summarize, for non-relativistic transients, or initially
relativistic off-axis transients, the characteristic timescale at
radio frequencies is generally determined by $t_{\rm dec}$.  However,
if the density is sufficiently high, such that $\nu_a(t_{\rm dec})
\gtrsim \nu_{\rm obs}$, the peak time will be longer, defined by time
at which $\nu_a=\nu_{\rm obs}$.  Similarly, if the source is
relativistic such that $\nu_m(t_{\rm dec})\gtrsim\nu_{\rm obs}$ the
peak time will also be longer than $t_{\rm dec}$, defined by the time
at which $\nu_m=\nu_{\rm obs}$. This means that for synchrotron
transients the timescale at a given luminosity cannot be made
arbitrarily short.  The boundary defining the allowed phase-space of
luminosity and timescale based on the above considerations is shown in
Figure~\ref{fig:Tc}.  This boundary is based on an initially
relativistic source decelerating into a constant density ambient
medium, and with $\nu_a\lesssim\nu_m$.  For any other case (e.g.,
non-relativistic source, $\nu_a\gtrsim \nu_m$) the timescale at a
given luminosity will be even longer.

Independent of the generic arguments above, synchrotron sources also
obey a basic physical constraint that restricts how rapidly they can
exhibit significant brightness variations, namely their brightness
temperature\footnotemark\footnotetext{We note that the variability
  timescales and brightness temperatures for various radio sources
  were recently compiled by \citet{Pietka+15}.} is $T_B\lesssim
10^{12}$ K (the so-called inverse Compton catastrophe limit;
\citealt{kpt69}).  The co-moving brightness temperature of an
expanding source at a time $t$ is given by (e.g.,
\citealt{Kulkarni+98})
\begin{equation}
  T_B' = \frac{L_{\nu}}{8\pi^{2} k}\frac{\Gamma^{2}}
  {\mathcal{D}\beta^{2}\nu^{2}t^{2}} \lesssim 10^{12}\,{\rm K},
\label{eq:TB}
\end{equation}
where $\mathcal{D} \approx \Gamma$ is the Doppler factor in cases when
the source is expanding towards the observer.  A source cannot vary on
a timescale shorter than its light crossing time, $\delta t_{\rm min}
\approx t\beta /2\Gamma^{2}$, where the $\sim 1/2\Gamma^{2}$ factor
corrects for photon arrival delay in the case of relativistic
expansion.  Equation~\ref{eq:TB} can thus be recast as a constrained
relation between the timescale and luminosity:
\begin{equation}
  \delta t_{\rm min} =\left[\frac{L_{\nu}}{16\pi^{2} kT_{\rm
        B}'\nu^{2}\Gamma^{3}}\right]^{1/2} \approx 80\,{\rm
    d}\,\,\Gamma^{-3/2} L_{\nu,30}^{1/2}\nu_{\rm GHz}^{-1},
\label{eq:FnuTC}
\end{equation}
where $L_{\nu,30} \equiv L_{\nu}/(10^{30}$ erg s$^{-1}$ Hz$^{-1}$) and the right-hand term takes $T_{\rm B}' = 10^{12}$ K. This leads
to the same conclusion that more luminous sources have longer
characteristic timescales.  We show the boundary based on the
condition $T_B\lesssim 10^{12}$ K for sources with $\Gamma\beta=1$ in
Figure~\ref{fig:Tc}.  This boundary is remarkably similar to the one
defined by the synchrotron model considerations discussed above.

Both considerations imply that mJy-level GHz transients, which are
roughly at the detection threshold of planned near-term surveys
(\autoref{tab:surveys}), will have characteristic durations of
$\gtrsim 100$ d at cosmological distances.  Shorter durations are
possible, but only if the luminosities are correspondingly lower,
making the detection volume much smaller for such sources; i.e., they
will only be detectable if they have a high volumetric rate.  For a
fixed transient luminosity, timescale, and volumetric rate, it is
straight-forward to compute the instantaneous number of events on the
sky above a given flux density threshold ($F_{\nu,\rm lim}$):
\begin{eqnarray}
  N_{\rm char} = \frac{4\pi}{3} \left(\frac{L_\nu}{4\pi F_{\nu,\rm
        lim}}\right)^{3/2} \mathcal{R} \, t_{\rm dur} \approx 0.9
  L_{\nu,30}^{3/2} \mathcal{R} t_{\rm dur,2},
  \label{eq:nchar}
\end{eqnarray}
where $\mathcal{R}$ is in units of Gpc$^{-3}$ yr$^{-1}$ and we use
$F_{\nu,\rm lim} = 1$ mJy.  In \autoref{fig:Tc} we show contours of
$\mathcal{R}$ such that one event could possibly be detected in a
three-year transient survey: given values for $L_\nu$ and $t_{\rm
  dur}$ we find $N_{\rm char} (\mathcal{R}) = 1$ in
\autoref{eq:nchar}, adopting $t_{\rm dur} = 3$ yr if the actual
transient duration is shorter than this value.  We find that even with
this generous definition of detectability, transients varying on a
timescale as short as $\sim 3$ days are only detectable if their
volumetric rate is $\gtrsim 10^{3}$ Gpc$^{-3}$ yr$^{-1}$.  This is
much higher than the rate of all known extragalactic transients with
the exception of Type Ib/c SNe (although the latter are
non-relativistic sources and thus vary on much longer timescales than
indicated by the condition $T_B\approx 10^{12}$ K).  For more
reasonable volumetric rates of relativistic transients ($\lesssim 1$
Gpc$^{-3}$ yr$^{-1}$; \autoref{tab:classes}) the minimum timescale is
$\sim 30$ days; this timescale increases to $\sim 100$ days for a
detection rate of a few events per year.  Thus, extragalactic
synchrotron radio transients with timescales of $\ll 100$ days are
unlikely to be detected in upcoming surveys.

\subsection{Models and Volumetric Rates}
\label{sec:models}

\begin{figure*}
\includegraphics[width=0.5\textwidth] {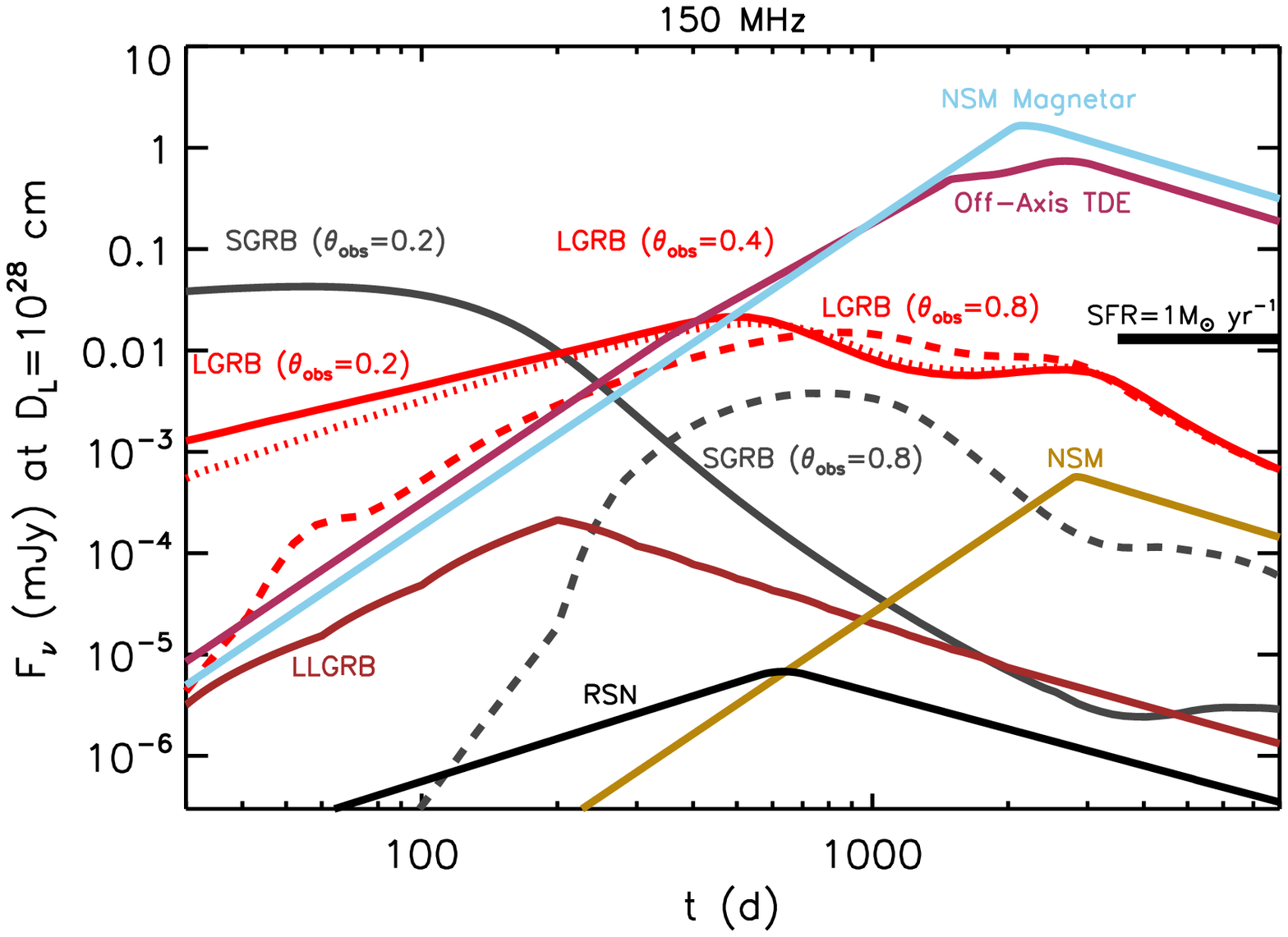}
\includegraphics[width=0.5\textwidth] {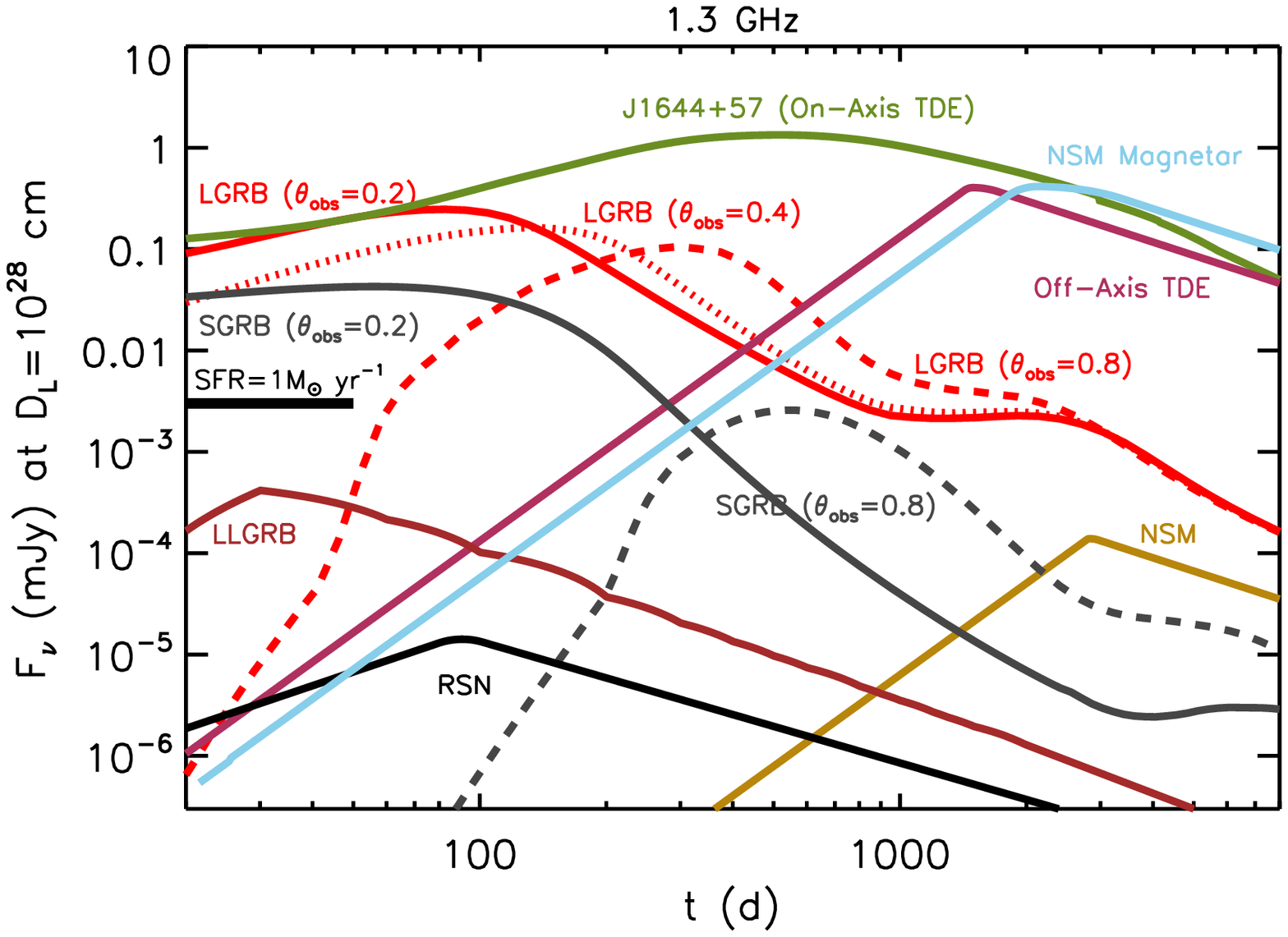}
\includegraphics[width=0.5\textwidth] {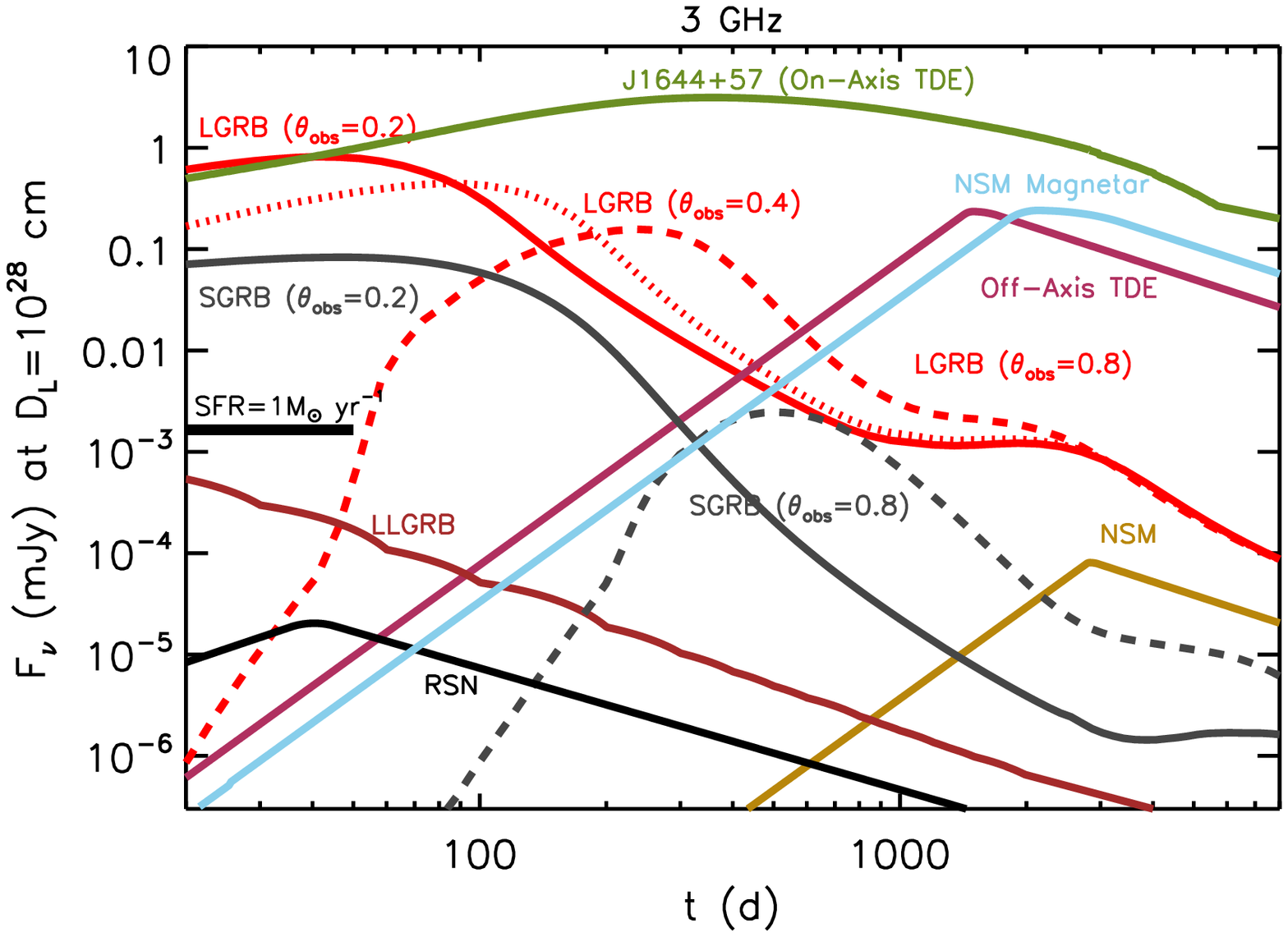}
\includegraphics[width=0.5\textwidth] {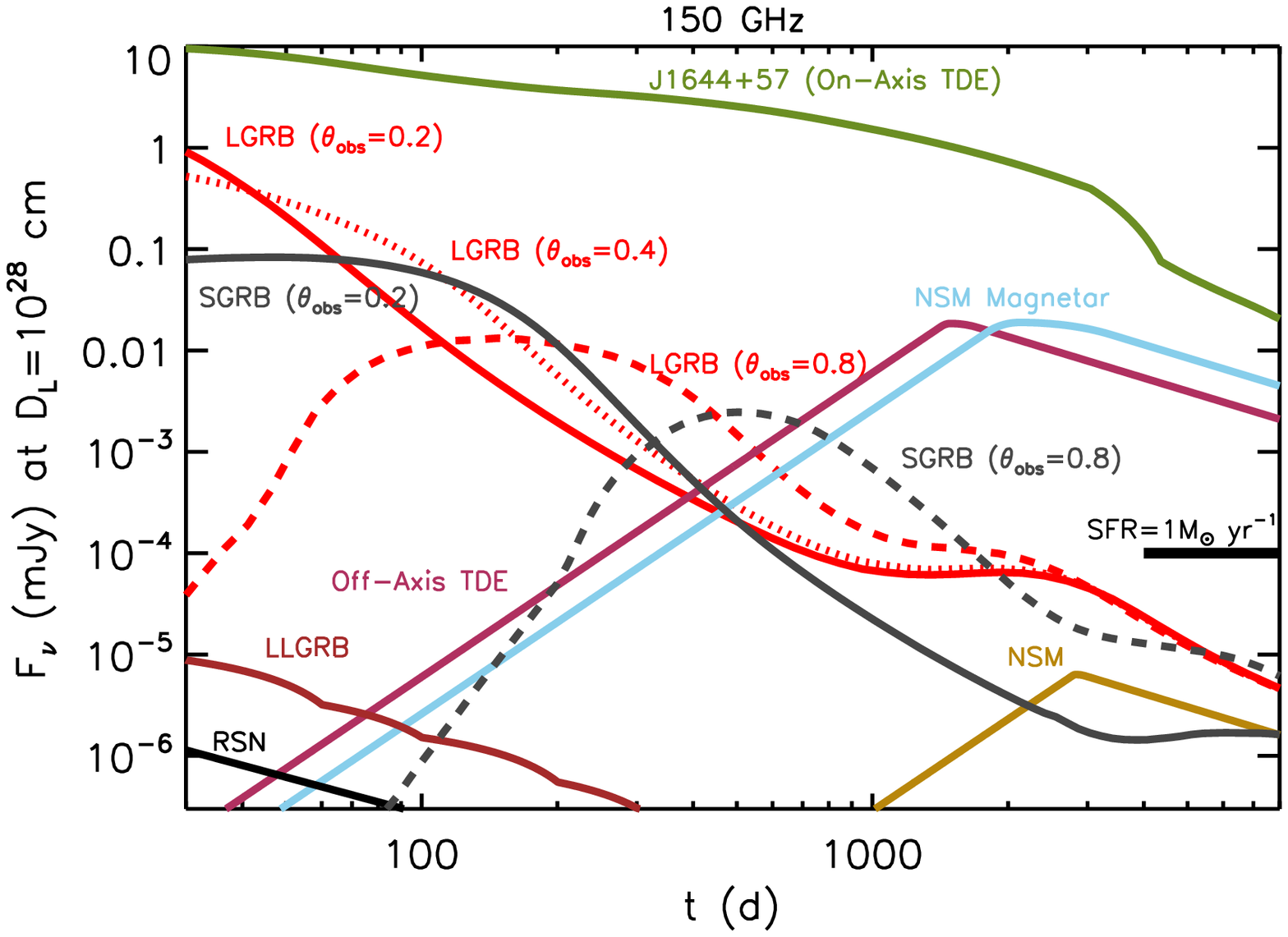}
\caption{Light curves of the various transients considered in this
  paper (Table~\ref{tab:classes}) at observing frequencies of $150$
  MHz (upper left), $1.3$ GHz (upper right), $3$ GHz (lower left), and
  $150$ GHz (lower right). Sources considered here include: on- and
  off-axis long GRB afterglows for various observing angles
  $\theta_{\rm obs}$ (LGRB: {\it red}); low-luminosity LGRB afterglows
  (LLGRB; {\it brown}); Type Ib/c SNe (RSN; {\it black}); on- and
  off-axis short GRB afterglows (SGRB: {\it charcoal}); on-axis jetted
  TDEs (\swsixteen: {\it green}); off-axis TDEs ({\it maroon}); neutron
  star binary mergers with prompt black hole formation (NSM; {\it
    tan}) and a magnetar remnant (NSM magnetar; {\it blue}). Flux
  densities are normalized to $D_{\rm L} = 10^{28}$ cm ($z\approx
  0.55$). In each panel we also annotate the typical flux density of a
  host galaxy forming stars at a fiducial rate of
  1~M$_\odot$~yr$^{-1}$ (\textit{horizontal black bars}). See
  Section~\ref{sec:classes} for further details.}
\label{fig:LCs}
\end{figure*}

In Figure~\ref{fig:LCs} we show the light curves for the various
transients that we consider in this paper at four frequencies (0.15,
1.3, 3, and 150 GHz) characteristic of planned and hypothetical
surveys (Table~\ref{tab:surveys} and \S\ref{sec:surveys}). As can be
seen in Figure~\ref{fig:Tc}, these transients span a wide range in the
luminosity-timescale phase-space and therefore provide a broad view of
extragalactic transients. It is also clear from Figure~\ref{fig:Tc}
that these transients indeed obey the boundaries defined by the basic
synchrotron model consideration discussed above and by the condition
$T_B\lesssim 10^{12}$ K.  Finally, by comparing the actual volumetric
rates of these transients (\autoref{tab:classes}) to the contours
showing the rates required for our fiducial level of observability
(one event brighter than $\sim 1$ mJy on the sky over three years), it
is already clear that few detections will be made in GHz surveys with
mJy sensitivity.

In the following sections we discuss each class of transients in
detail. Unless otherwise noted we adopt the microphysical shock
parameters $\epsilon_e = 0.1$, $\epsilon_{B}= 0.01$, and $p=2.3$,
typical of those found or used for known extragalactic transients.

\subsubsection{Classical Long GRBs}
\label{sec:LGRBs}

The afterglows of long GRBs (LGRBs) are powered by the deceleration of
an ultra-relativistic jet by interaction with the circumburst medium
(\citealt{Meszaros&Rees97}). The isotropic equivalent energies of
classical LGRBs are large, $E_{\rm iso} \approx 10^{53}-10^{54}$ erg,
but with typical jet opening angles of $\theta_j \sim 0.1$
\citep{Bloom+03} the beaming-corrected energies are $\sim
10^{50}-10^{52}$ erg (e.g., \citealt{Frail+01,Berger+03,Cenko+12}).
This has been confirmed by late-time radio calorimetry
\citep{Frail+00,Berger+04,Frail+05,Shivvers&Berger11}. LGRB jets
interact with either the interstellar medium or the wind of the
massive progenitor star, with typical values of $n\sim 0.1-10$
cm$^{-3}$.

Here we consider synchrotron emission from the forward shock,
neglecting emission from the reverse shock which travels into the
ejecta at early time (e.g., \citealt{Sari&Piran99}). This is justified
by the fact that the reverse shock emission is generally fainter than
the forward shock emission at the relevant frequencies of $\lesssim
{\rm GHz}$ (e.g., \citealt{Kulkarni+99b,Laskar+13}).  However, at the
high frequencies relevant to millimeter surveys, the reverse shock
emission can be substantially brighter than the forward shock for
$\sim {\rm day}$ \citep{Laskar+13}; we discuss the implications of
this point in \S\ref{sec:discussion}.

The observed light curves of LGRB radio afterglows also depend on the observer
viewing angle with respect to the jet axis, $\theta_{\rm obs}$. We thus
consider both on- and off-axis LGRB afterglow light curves, based on the
relativistic numerical hydrodynamical simulations of
\citet{VanEerten+10}\footnotemark\footnotetext{\url{http://cosmo.nyu.edu/afterglowlibrary}},
including self-absorption. We utilize a model with properties typical of
observed LGRBs: $\theta_j=\theta_{\rm obs}=0.2$, $E_K = 10^{51}$ erg, $n=1$
cm$^{-3}$, and $p= 2.5$, as well as off-axis (``orphan'') jets with
$\theta_{\rm obs} = 0.4,\,\, 0.8,\,\, 1.57$ (Figure~\ref{fig:LCs}). On-axis
events, though more luminous, represent only a small fraction, $f_{\rm obs} =
\int_{\theta_{1}}^{\theta_{2}}\sin \theta d\theta$, of the total event rate;
the integration from $\theta_1 = \theta_{\rm obs}/2$ to $\theta_2 =
\theta_{\rm obs}$ covers the approximate range of solid angle covered by
observers with viewing angle $\theta \approx \theta_{\rm obs}$. Depending on
observing frequency, depth, and cadence of the survey, detection rates may be
dominated by on- or off-axis events.

For the local volumetric rate of on-axis LGRBs we adopt a fiducial
value of $\mathcal{R}_{\rm LGRB}(\theta_{\rm obs} \lesssim \theta_{\rm
  j}) = 0.3$ Gpc$^{-3}$ yr$^{-1}$, motivated as follows.
\citet{Wanderman&Piran10} estimate $\mathcal{R}_{\rm LGRB}\sim 1$
Gpc$^{-3}$ yr$^{-1}$ for events with isotropic luminosity of $L_{\rm
  iso} \gtrsim 10^{50}$ erg s$^{-1}$ ($E_{\rm iso}\sim 10^{51}$ erg
for a typical LGRB duration of 10 seconds). The LGRB luminosity
function measured by {\it Swift} is approximately flat per unit log
$L_{\rm iso}$ below the characteristic value $L_{\rm iso} \approx
10^{52}$ erg s$^{-1}$ (corresponding to the same isotropic energy
$\sim 10^{53}$ erg in our assumed afterglow calculation above).  Thus
the rate of events with $E_{\rm iso} \sim 10^{53}$ erg is $\sim 1/3$
of the rate of events with $E_{\rm iso} \gtrsim 10^{51}$ ergs, or
$\approx 0.3$ Gpc$^{-3}$ yr$^{-1}$.  For off-axis LGRBs we scale the
on-axis rate according to $\mathcal{R}_{\rm LGRB}(\theta_{\rm obs}) =
\mathcal{R}_{\rm LGRB}(\theta_{\rm obs} = \theta_{\rm j})f_{\rm
  obs}^{-1}$.  Finally, the rate is assumed to evolve with redshift in
the same manner as the cosmic star formation rate density, according
to the model of \citet{Cucciati+12}.  Although this may not hold
strictly (e.g., \citealt{Yuksel+08}), the uncertainty introduced by
this assumption is small compared to that from other approximations.

\subsubsection{Low-Luminosity Long GRBs}

A class of low-luminosity LGRBs (LLGRBs) with lower total
beaming-corrected energies of $E_K\sim 10^{49}$ erg and only
mildly-relativistic ejecta ($\Gamma_i \approx 2$) has been identified
in recent years \citep{Kulkarni+98,Soderberg+04}.  LLGRBs appear to be
roughly isotropic, and it has been suggested that they result from
relativistic shock break-out following the core-collapse of massive
stars with extended envelopes (e.g., \citealt{Matzner&McKee99},
\citealt{Nakar&Sari12}).  We model the radio emission from LLGRBs
using the model of \citet{BarniolDuran+14} as fit to the radio
afterglow of GRB\,980425 \citep{Kulkarni+98}.

We use an LLGRB volumetric rate that is 10 times higher than the
beaming-corrected LGRB rate (e.g.,
\citealt{Soderberg+04,Coward05,Guetta&DellaValle07}).  Adopting a
total on-axis LGRB rate of $\approx 1$ Gpc$^{-3}$ yr$^{-1}$ and a
classical LGRB beaming fraction of $\approx 50$ we thus take the
volumetric rate of LLGRBs to be $\mathcal{R}_{\rm LLGRB} \approx 500$
Gpc$^{-3}$ yr$^{-1}$.  We note that given their lower energy scale,
LLGRBs are only detectable nearby, and therefore redshift evolution of
the rate is unimportant.

\subsubsection{Short GRBs}

Short-duration gamma-ray bursts (SGRBs) are also accompanied by
afterglow emission, but with beaming-corrected energies of $E_K\sim
10^{49}-10^{50}$ erg, lower than in the case of LGRBs
\citep{Berger14}.  The circumburst densities are also theoretically
expected and observationally inferred to be lower than in the case of
LGRBs, with $n\sim 0.001-0.1$ cm$^{-3}$ (e.g., \citealt{Fong+13}).  We
therefore adopt the low energy afterglow model of
\citet{VanEerten&MacFadyen11} with $E_K=10^{50}$ erg, $n=10^{-3}$
cm$^{-1}$, $\theta_j=0.2$, $\epsilon_e = \epsilon_B = 0.1$, and $p =
2.5$.  These parameters are consistent with the observed on-axis
optical afterglows of SGRBs \citep{Metzger&Berger12}.  In addition to
the on-axis case ($\theta_{\rm obs}=0.2$) we also consider off-axis
events with $\theta_{\rm obs} = 0.4,\,\, 0.8,\,\, 1.57$.

We use a local volumetric rate of on-axis SGRBs of $\mathcal{R}_{\rm
  SGRB} = 5$ Gpc$^{-3}$ yr$^{-1}$ (e.g., \citealt{Wanderman&Piran14}),
and scale it appropriately for off-axis angles as described in
\S\ref{sec:LGRBs}.  As in the case of LLGRBs, redshift evolution of
the rate is unimportant due to the low luminosity of these events.

\subsubsection{NS-NS Mergers with Prompt Black Hole Formation}

The merger of two neutron stars, or of a neutron star and a low mass
black hole, results in the ejection of a small quantity of mass,
$M_{\rm ej} \sim 10^{-3}-10^{-2}M_{\odot}$, with mildly relativistic
velocities, $\beta_i = 0.1-0.3$ (e.g.,
\citealt{Rosswog+13,Hotokezaka+13}).  The interaction of the ejecta
with the surrounding interstellar medium will result in synchrotron
radio emission similar to that of a supernova remnant
\citep{Nakar&Piran11,Hotokezaka&Piran15}.  We calculate the radio
emission using the spherical model of \citet{Nakar&Piran11} for
$\beta_i=0.2$, $E_K=3\times 10^{50}$ erg ($M_{\rm ej} =
10^{-2}M_{\odot}$), and $n = 0.1$ cm$^{-3}$.

We use a local neutron star merger (NSM) volumetric rate of
$\mathcal{R}_{\rm NSM} = 500$ Gpc$^{-3}$ yr$^{-1}$, consistent with
that inferred from the Galactic binary pulsar population (e.g.,
\citealt{Kim+13}).

\subsubsection{NS-NS Mergers Leaving a Stable NS Remnant \& White
  Dwarf Accretion-Induced Collapse}
\label{sec:magnetar}

In some cases neutron star binary mergers may result in the formation
of a long-lived neutron star instead of immediate collapse to a black
hole (e.g., \citealt{Metzger+08,Siegel+14}).  This stable remnant is
formed rapidly rotating, with a period of $P\sim 1$ ms, and
correspondingly large rotational kinetic energy of $E_{\rm rot}
\approx 4\pi^{2}I/P^{2}\approx 3\times 10^{52}$ erg, where $I\sim
10^{45}$ g cm$^{2}$ is the NS moment of inertia.  If the merger
remnant also possesses a moderately strong dipole magnetic field of
$B\gtrsim 10^{13}$ G, then its electromagnetic dipole spin-down will
transfer the rotational energy to the small quantity of ejecta,
accelerating it to trans-relativistic speeds, $\Gamma_i\simeq E_{\rm
  rot}/M_{\rm ej}c^{2}\gtrsim 1$, before the ejecta have been
decelerated by the ISM \citep{Metzger&Bower14}.  A millisecond neutron
star with similar properties may also be formed by the
accretion-induced collapse (AIC) of a rotating white dwarf
(\citealt{Usov92}).

We model the radio emission from both neutron star mergers leaving a
stable remnant and AIC events using the \citet{Nakar&Piran11}
spherical blast wave
model\footnotemark\footnotetext{\citet{Piro&Kulkarni13} argue that the
  nascent pulsar wind nebula created by AIC will also produce a radio
  transient, but we estimate that the synchrotron radiation from the
  ISM interaction will dominate the nebular emission unless the ISM
  density is exceedingly low, $n\ll 10^{-4}\epsilon_{e,-1}^{-1}$
  cm$^{-3}$.} with $\beta_i = 1$, $E_K=3\times 10^{52}$ erg, and
$n=0.1$ cm$^{-3}$.  The energy and velocity of the ejecta are
uncertain because they depend, among other things, on how efficiently
the NS outflow couples its energy to the merger ejecta (e.g.,
\citealt{Bucciantini+12}) and whether a fraction of the NS rotational
energy is instead lost to gravitational radiation (e.g.,
\citealt{Corsi&Meszaros09}).  The energy will also be smaller if the
NS collapses to a black hole before the bulk of its rotational energy
is extracted; however, the much brighter radio emission from stable NS
remnants imply that they will dominate the radio detection rate if
such objects form at all.  A higher ISM density is expected compared to
the case of prompt BH formation because mergers leaving stable
remnants may occur preferentially within their host galaxies due to
the small natal kicks that are expected to accompany the birth of the
lowest mass neutron stars formed in electron capture supernovae (e.g.,
\citealt{Belczynski+08}).

The rate of mergers leaving stable NS remnants is unknown, as it
depends on the equation of state of ultra-high density matter and on
the mass distribution of binary neutron stars.  We therefore scale the
rate to $1\%$ of the total NSM rate, i.e., $\sim 5$ Gpc$^{-3}$
yr$^{-1}$; as we show later on, much higher volumetric rates may
already be ruled out by past radio surveys due to the extremely high
luminosities of these events (see in particular the FIRST/NVSS
constraints from \citet{Levinson+02} in Figure~\ref{fig:lnls}). The
rate of AIC is also uncertain, with population synthesis estimates
spanning $\mathcal{R}_{\rm AIC}\sim 10-10^{3}$ Gpc$^{-3}$ yr$^{-1}$
\citep{Yungelson&Livio98}, and the fraction of WDs that are rapidly
rotating at the time of collapse is even less certain.  We scale the
volumetric rates with redshift according to the cosmic star formation
rate density (\S\ref{sec:LGRBs}), assuming that the characteristic
delay time between star formation and these events is relatively
short, $\ll$ Gyr (e.g., \citealt{Leibler&Berger10}). We note that if
the typical delays are in fact much longer, the observed rate will be
suppressed due to the loss of events at $z\gtrsim 1$.  Given that both
NS mergers with stable magnetar remnants and rotating AIC events are
predicted to produce similar radio transients and that the rates of
both are highly uncertain, for notational simplicity we hereafter
denote both models as ``NSM-magnetar.''

\subsubsection{On-Axis Jetted Tidal Disruption Event (\swsixteen)}

The transient event {\it Swift} J$164449.3+573451$ (hereafter
\swsixteen) exhibited several properties that led to its
interpretation as a tidal disruption event with a relativistic jet
\citep{Bloom+11,Burrows+11,Levan+11,Zauderer+11}.  The relativistic
outflow produced long-lived radio synchrotron emission, which is still
detected at the present
\citep{Giannios&Metzger11,Zauderer+11,Berger+12,Metzger+12,Zauderer+13}.
A second TDE candidate with similar X-ray and radio emission, Sw
J2058+05, has been reported by \citet{Cenko+12}.

We model the radio emission from on-axis jetted TDEs at $\gtrsim 1$
GHz directly using the radio light curves of \swsixteen\ as a template
\citep{Zauderer+11,Berger+12,Zauderer+13}.  Since data are not yet
available at $150$ MHz \citep{Cendes+14}, we assume that the on-axis
emission at this frequency is identical to the isotropic, off-axis
model (see below).  This is justified by the high self-absorption
frequency at early times, which will strongly suppress the low
frequency emission for on-axis sources.

We estimate the local volumetric rate of on-axis jetted TDEs in two
ways.  First, the fact that one event was detected by {\it Swift} in
$\Delta t\approx 10$ years of monitoring to a redshift $z = 0.35$
(co-moving volume of $V\approx 11$ Gpc$^{-3}$) suggests a rate of
$\mathcal{R}_{\rm Sw J1644+57}\sim 1/(V\Delta T) \sim 0.01$ Gpc$^{-3}$
yr$^{-1}$.  Alternatively, we can utilize the theoretically expected
TDE rate of $\sim 10^{-5}-10^{-4}$ galaxy$^{-1}$ yr$^{-1}$ (e.g.,
\citealt{Wang&Merritt04,Stone&Metzger15}), which given a local galaxy
density of $\sim 10^{-2}$ Mpc$^{-3}$ corresponds to a TDE rate of
$\sim 100-1000$ Gpc$^{-3}$ yr$^{-1}$.  Assuming a beaming correction
of $\sim 100$ for \swsixteen\ \citep{Zauderer+11,Berger+12,Metzger+12}
and that $\lesssim 10\%$ of TDEs produce relativistic jets
\citep{Bower+13,vanVelzen+13}, this corresponds to a volumetric rate
of $\lesssim 0.1-1$ Gpc$^{-3}$ yr$^{-1}$ for on-axis jetted TDEs,
consistent with the empirical estimate. In what follows we adopt the
conservative on-axis rate of $\mathcal{R}_{\rm Sw J1644+57}\sim 0.01$
Gpc$^{-3}$ yr$^{-1}$ but note that the true local rate may be higher
by at least an order of magnitude, especially when considering that
the conditions to produce a mildly-relativistic outflow with
$\Gamma_i\gtrsim 2$, which will produce radio emission, may be less
stringent than those required to produce bright X-ray emission (and
hence that \swsixteen\ provides only a lower limit on the rate).

We scale the TDE rate with redshift using the volume density of $\sim
10^{5}-10^{7}M_{\odot}$ black holes, for which we employ the model of
\citet{Sijacki+14} (their Figure~2).  This model predicts that the TDE
rate decreases by a factor of $\sim 2$ between $z = 0$ and $z\sim 4$,
before decreasing more rapidly at $z\gtrsim 4$ (although see
\citealt{Hopkins+07} who find a more rapid evolution with redshift).
Beyond the uncertainties in the evolution of the supermassive black
hole volume density, the TDE rate per galaxy also depends on other
factors, such as the mass distribution of stars in galactic nuclei,
which could also evolve strongly with redshift.  Our assumed rate
evolution thus represents a best guess given the lack of detailed
studies.

\subsubsection{Off-Axis Jetted Tidal Disruption Events}

For \swsixteen\ the initial jet Lorentz factor was estimated to be
$\Gamma_i \sim 10$ \citep{Metzger+12,Berger+12}, corresponding to a
beaming fraction $f_{\rm b} \sim 0.01$.  Off-axis jetted TDEs may thus
dominate on-axis events at radio frequencies
\citep{Giannios&Metzger11,vanVelzen+11,Donnarumma+15}.  Detailed
predictions for the off-axis radio light curves of jetted TDEs are not
currently available, but because the emission is expected to be
relatively isotropic and only mildly-relativistic at late times (e.g.,
\citealt{Berger+12}), we assume a spherical blast wave model.  We
utilize the \citet{Nakar&Piran11} model with $\beta_i = 1$,
$E_K=10^{52}$ erg, $n=0.1$ cm$^{-3}$, $\epsilon_e = 0.2$,
$\epsilon_{\rm B} = 0.01$, and $p = 2.3$, chosen to produce a 1.4 GHz
peak flux density in agreement with that of \swsixteen\ (since at late
times the jet has spread laterally and its emission is approximately
isotropic).

We assume the same evolution of the volumetric rate with redshift as
in the on-axis case, but with a local volumetric rate of $\sim 1$
Gpc$^{-3}$ yr$^{-1}$ which is $f_{\rm b}^{-1}\sim 100$ times the
on-axis rate.

\subsubsection{Type Ib/c Supernovae}
\label{sec:RSNe}


Core-collapse supernovae (SNe) produce radio emission as the fastest
ejecta interact with the dense gas of the progenitor's stellar wind
(e.g., \citealt{Chevalier82}).  Radio emission is observed in both
stripped-envelope Type Ib/c SNe (e.g., \citealt{Berger+03}) and Type
II SNe (e.g., \citealt{Weiler+02}).  In fact, the few genuine radio
transient discoveries to date have been identified as Type II SNe: a
SN in Markarian 297 \citep{Yin&Heeschen91}, \object{FIRST
  J121550.2+130654} \citep{Levinson+02, Gal-Yam+06}, and \object{SN
  2008iz} \citep{bmr+09, Brunthaler+10}.  We note, however, that only
\object{FIRST J121550.2+130654} was discovered in an untargeted search
(using a comparison of NVSS and FIRST with a baseline of about 5
years), while the other two events were discovered in targeted
observations of startburst galaxies with elevated core-collapse SN
rates.  Theory and observations indicate that Type II SNe are actually
challenging to discover in the untargeted surveys we consider here.
In particular, these sources are dominated by synchrotron
self-absorption and free-free absorption such that the most luminous
events (generally Type IIn SNe, with $L_\nu\sim 10^{28}-10^{29}$ erg
s$^{-1}$ Hz$^{-1}$) have typical timescales of about a decade at $\sim
1$ GHz \citep{Chevalier06}, and are thus indistinguishable from steady
sources; Type II SNe that vary on timescales of $\sim {\rm year}$
(generally Type IIP SNe) have much lower luminosities ($L_\nu\lesssim
10^{27}$ erg s$^{-1}$ Hz$^{-1}$) and are thus detectable in a
negligible volume.  This indicates that Type II SNe may appear in the
source lists of radio surveys, but will not be easily identified as
transients within the time-frame of the surveys.

Here we instead focus on Type Ib/c SNe, for which the radio emission
evolves more rapidly due to the presence of faster ejecta ($\beta\sim
0.1-0.3$; \citealt{Berger+03,Chevalier&Fransson06}). The Type Ib/c SNe
detected in the radio exhibit a wide spread in peak luminosity and
timescale ($L_{\nu,p}\sim 10^{25}-10^{28}$ erg s$^{-1}$ Hz$^{-1}$ and
$t_p\sim 10-1000$ d), following the general pattern in which the most
luminous events evolve most slowly.  At the present, the radio
luminosity function of Type Ib/c SNe has not been quantified so here
we use data for the well-sampled, and relatively typical, event
SN\,2008D \citep{Soderberg+08}, modeled using the formulation of
\citet{Soderberg+05}.  We note that the SN model is somewhat different
from the models we consider above because the ejecta have a large
spread in velocity, with the energy distributed approximately as
$E_K({>}v) \sim v^{-5}$ \citep{Chevalier82}.

To estimate the local volumetric rate we note that the fraction of
Type Ib/c SNe with detectable radio emission approaches unity only for
the nearest events ($\lesssim 15$ Mpc).  About $\sim 1/4$ of events
are detected within the distance of SN\,2008D ($\approx 27$ Mpc).  We
therefore assume a fiducial rate of $\mathcal{R}_{\rm Ibc} = 5\times
10^{3}$ Gpc$^{-3}$ yr$^{-1}$, corresponding to one quarter of the Type
Ib/c SN volumetric rate \citep{Guetta&DellaValle07}.

\section{Radio Surveys}
\label{sec:surveys}

A variety of time-domain radio surveys covering a broad range of
frequencies, sensitivities, cadences, and angular resolutions are
slated to come online over the course of the next decade and
beyond. This section summarizes the set of surveys that we simulate in
this work.  Some are currently in advanced stages of development or
are even already taking data, while others are still in early stages
of planning.  We also consider hypothetical surveys, some of which
depend on judicious temporal binning of planned high-cadence surveys,
motivated by the transient timescale arguments we presented in
\S\ref{sec:general}.  The key parameters of the various surveys are
summarized in \autoref{tab:surveys}.

\begin{deluxetable*}{lccccccccc}
\tabletypesize{\scriptsize}
\tablecolumns{8}
\tabcolsep0.05in\footnotesize
\tablewidth{0pc}
\tablecaption{Adopted Radio Survey Parameters
\label{tab:surveys}}
\tablehead {
\colhead{Name}                   &
\colhead{$\theta_{\rm res}$}  &
\colhead{$\nu$}                  &
\colhead{$\sigma^{(a)}$}      &
\colhead{Area}                      &
\colhead{$\Delta T$$^{(b)}$} &
\colhead{N$_{\rm obs}$$^{(c)}$} &
\colhead{Ref.}                        \\
\colhead{}              &
\colhead{(arcsec)}  &
\colhead{(GHz)}      &
\colhead{(mJy)}       &
\colhead{(deg$^{2}$)}          &
\colhead{(d)}           &
\colhead{}
}
\startdata
LOFAR & 80 & 0.15 & 12 & 1500 & 30 & 36 & 1\\
LOFAR-Expanded  & 80 & 0.15 & 1.2 & $10^{4}$  & 30  & 36 & 1\\
SKA-Low & 11 & 0.15 & 0.004 & $10^{4}$ & 90 & 13 & 3 \\
\hline
VAST-Wide  & 10 & 1.3 & 0.5 & 10$^{4}$ & 1   & 1096 & 2\\
VAST-Wide-Stack  & 10 & 1.3 & 0.09 & 10$^{4}$ & 30 & 36 & 2\\
VAST-Deep & 10 & 1.3 & 0.05 & 10$^{4}$  & 365  & 4 & 2 \\
VAST-Deep-SF   & 10 & 1.3 & 0.05 & 30  & 1  & 365 & 2 \\
VAST-Deep-SF-Stack   & 10 & 1.3 & 0.01 & 30  & 30  & 36 & 2 \\
SKA & 0.9 & 1.3 & 0.009 & 10$^{4}$  & 90  & 13 & 3 \\
SKA-Expanded & 0.9 & 1.3 & 0.006 & $3\times 10^{4}$  & 30  & 36 & 3 \\
\hline
VLASS-Wide  & 2 & 3 & 0.1 & $10^{4}$ & 365  & 4 & - \\
VLASS-Deep  & 0.6 & 3 & 0.003 & 10  & 365  & 4 & - \\
\hline
CMB  & 840 & 150 & 1.0 & 10$^{4}$  & 1  & 1096 &  4\\
CMB-Stack-1  & 840 & 150 & 0.31 & 10$^{4}$  & 10  & 110 & 4 \\
CMB-Stack-2  & 840 & 150 & 0.1 & 10$^{4}$  & 100  & 11 & 4
\enddata
\tablecomments{All surveys are assumed to last 3 years, except for
  VAST-Deep-SF with a duration of 1 year. $^{(a)}$RMS $1\sigma$
  sensitivity; $^{(b)}$Cadence; $^{(c)}$Number of epochs.  References:
  (1) \citealt{Fender+12}, J. Broderick (private communication); (2)
  \citealt{Murphy+13}; (3) \citealt{cr04}; (4) B.~Johnson, G.~Jones,
  private communication.}
\end{deluxetable*}

\subsection{ASKAP VAST}
\label{sec:ASKAP}

The Australian Square Kilometer Array Pathfinder (ASKAP) is a
precursor and technology development platform for the SKA, under
development in Western Australia.  Its wide field of view ($\sim 30$
deg$^{2}$) and moderately high sensitivity ($\sim {\rm mJy}$
beam$^{-1}$ for a 10 s integration) will enable fast, sensitive
all-sky surveys \citep{Johnston+08}. The ASKAP Survey for Variables
and Slow Transients (VAST; \citealt{Murphy+13}) comprises several
surveys aimed at detecting transients in the $1.1-1.4$ GHz frequency
range: `VAST-Wide' will survey $10^4$ deg$^{2}$ per day to a $1\sigma$
rms sensitivity of $0.5$ mJy, while VAST-Deep Multi-Field (hereafter
`VAST-Deep'), will survey $10^4$ deg$^{2}$ to $0.05$ mJy ($1\sigma$)
every $\sim {\rm year}$.  We also consider the planned Deep Single
Field survey (hereafter `VAST-Deep-SF') with $30$ deg$^{2}$ surveyed
daily to $0.05$ mJy ($1\sigma$).

We assume that both VAST-Wide and
VAST-Deep\footnotemark\footnotetext{The terms `Deep' and `Wide' are
  somewhat misleading; both surveys cover the same sky area, with the
  real distinction being the higher cadence and lower per-epoch
  sensitivity of VAST-Wide.  We nevertheless adopt this terminology
  here to avoid confusion with existing literature.} will last for 3
years, while VAST-Deep-SF is assumed to last for 1 year.  Although a
dynamical cadence has been proposed for VAST-Deep, we assume a uniform
cadence since details of the dynamical cadence have not yet been
specified.

For the daily-cadence surveys VAST-Wide and VAST-Deep-SF we also
consider the possibility of co-adding the daily exposures to create
deeper maps with a slower effective cadence, motivated by the long
durations of extragalactic synchrotron transient sources.  We
construct two separate mock surveys (`VAST-Wide-Stack' and
`VAST-Deep-SF-Stack') by averaging 30 separate daily observations of
VAST-Wide and VAST-Deep-SF to achieve effective sensitivities of
$0.09$ mJy and $0.01$ mJy, respectively.  We note that the latter is
limited to the estimated confusion limit of ASKAP.

\subsection{LOFAR Radio Sky Monitor}
\label{sec:LOFAR}

The LOw-Frequency ARrray (LOFAR) operates between about 30 and 240 MHz
\citep{VanHaarlem+13}.  The 32 core stations of LOFAR will ultimately
be able to return up to 24 individual beams, which will cover a patch
of the sky as a Radio Sky Monitor (RSM;
\citealt{Fender+06,Best+08,Fender+12}).  The currently operational
LOFAR RSM Zenith Monitoring Program covers $\approx 1500$ deg$^{2}$ at
$\approx 150$ MHz to a $1\sigma$ rms sensitivity of $12$ mJy
(J.~Broderick, private communication).  A mean cadence timescale of
$\sim 1$ month is planned, corresponding to 36 observing epochs for an
assumed three year survey.  Beyond the current plan, we also consider
a more ambitious hypothetical survey (`LOFAR-Expanded'), which we
assume reaches a factor of 10 times deeper and covers a significantly
larger sky area ($\sim 10^4$ deg$^{2}$).  Though ambitious in relation
to LOFAR's existing capabilities, such a program is similar to the
original RSM.

\subsection{VLA Sky Survey}
\label{sec:EVLA}

A series of large surveys with the upgraded Karl G. Jansky Very Large
Array \citep{Perley+11}, collectively known as the VLA Sky Survey
(VLASS), has been recently discussed.  Although the parameters of the
VLASS components are far from settled, we simulate several large
surveys using the most recent information available to us.  We
consider surveys operating in the S-band ($2-4$ GHz) using a
combination of the high-resolution A and B configurations.  The
`VLASS-Wide' survey covers $10^4$ deg$^2$ at an rms sensitivity of
$0.1$ mJy ($1\sigma$) over 4 epochs with a yearly cadence.  The
`VLASS-Deep' survey covers a much smaller area of $10$ deg$^{2}$ with
an annual cadence, achieving a $1\sigma$ sensitivity of $\sim 3\mu$Jy.
Although it is planned for each epoch of VLASS-Deep to achieve the
target sensitivity by revisiting the survey footprint in 10--30 passes
over $\sim 3$ months, we consider only the final deep integrations
because the timescale for multiple passes is comparable to that of
extragalactic synchrotron transients.  Finally, a VLASS `All-Sky'
survey covering about $34,000$ deg$^2$ with a sensitivity of $0.1$ mJy
($1\sigma$) is also being discussed, but it will be comprised of only
2 epochs, limiting its utility for detailed time-domain studies.  We
do not consider this survey here.

\subsection{Millimeter Survey}
\label{sec:CMB}

In recent years mm and sub-mm telescopes have been developed to study
the cosmic microwave background (CMB), which are sensitive,
wide-field, and exceptionally stable.  We consider whether these
instruments could be employed for transient discovery.  We estimate
what a hypothetical ground-based mm survey could achieve by
extrapolating the parameters of the proposed design for a polarimeter
based on lumped-element kinetic inductance detectors (LEKIDs;
\citealt{Araujo+14}).  This design specifies an aperture diameter of
50 cm, with an estimated $\sim 3.2$ mJy sensitivity ($1\sigma$), and a
daily sky coverage of $\sim 10^4$ deg$^{2}$ (G.~Jones, B.~Johnson,
private communication).  For the purpose of our study we consider a
large 2-m diameter instrument, achieving $1$ mJy sensitivity
($1\sigma$) with a daily cadence.  As in the case of ASKAP, we also
consider time-averaged versions of the data reaching $1\sigma$ rms
sensitivities of $0.3$ and $0.1$ mJy for cadences of 10 days
(`CMB-Stack-1') and 100 days (`CMB-Stack-2'), respectively.

One significant drawback of the wide-field CMB instruments is that
they have poor angular resolution ($14$ arcmin for the instrument
considered here).  This limits both the ability to identify
non-varying point sources due to confusion (e.g., \citealt{bsi+02})
and the quality of event localization.  However, the excellent
stability of CMB telescopes allows the detection of \textit{variable}
sources far below the confusion limit, and since synchrotron
transients evolve from high to low frequencies with time, follow-up of
mm transients with other facilities at cm wavelengths could lead to
better localizations.

\subsection{Square Kilometer Array}
\label{sec:ska}

The planned Square Kilometer Array (SKA) will be the most powerful
radio telescope ever built \citep{cr04}. The observatory will be
located in both Australia and South Africa and will initially be
composed of arrays of three distinct telescope designs denoted
SKA1-low, SKA1-mid, and SKA1-survey.  We consider two hypothetical
surveys for transients that could be conducted with the SKA1-survey
array, which is planned to have a field of view of $18$ deg$^{2}$ and
to operate in a frequency range of $0.65-1.67$ GHz.  The first (`SKA')
assumes that 100 hours will be spent every quarter in a survey of
$10^4$ deg$^{2}$, resulting in a $1\sigma$ rms sensitivity of $9$
$\mu$Jy.  The second (`SKA-Expanded') assumes that the SKA1-survey
array will be fully dedicated to a continuous transient survey of
$3\times 10^4$ deg$^{2}$ with a monthly cadence, resulting in a
sensitivity of $6$ $\mu$Jy ($1\sigma$).  This hypothetical survey
defines a practical upper limit to what untargeted GHz transient
surveys may be expected to yield in the coming decades. Finally, we
consider a low-frequency survey at $150$ MHz (`SKA-Low') assuming
observations of $10^4$ deg$^{2}$ with 100 hours per epoch on a
quarterly cadence, resulting in a sensitivity of $4$ $\mu$Jy
($1\sigma$).

\section{Simulated Transient Surveys}
\label{sec:monte}

In this section we combine the radio light curves and volumetric rates
of the transient sources described in \S\ref{sec:classes} with the
survey parameters defined in \S\ref{sec:surveys} to assess the
detection rates of various transients, as well as some of their basic
properties (e.g., redshift distribution), using Monte Carlo
simulations.

\subsection{Detection Criteria}
\label{sec:criterion}

Our basic criteria for the detection and identification of a transient
source are straightforward: (1) a $\gtrsim 10\sigma$ detection during
at least one observing epoch; and (2) at least a factor of 2 change in
brightness during the course of the survey (detection or $5\sigma$
upper limit).  The first criterion is motivated by the significant
potential for spurious signals given the experiences of untargeted
radio transient surveys to date \citep{Gal-Yam+06, Ofek+10, cbk+11,
  Frail+12}, as well as by the large number of beams and epochs
comprising each survey (e.g., $\sim 10^{9}-10^{11}$ beams per epoch
for the ASKAP, VLA, and SKA surveys). The second criterion effectively
defines the distinction between sources that can be identified as
genuine transients during the survey, and those that evolve too slowly
to be distinguished from steady or mildly variable sources. Indeed,
some variable radio sources can vary by about a factor of two on slow
timescales potentially leading to contamination if a robust
variability criterion, such as the one we use here, is not utilized
\citep{Becker+10,Ofek+11}.

We note that some previous works on the detectability of radio
transients have utilized $5\sigma$ as a fiducial detection threshold
(e.g., \citealt{cif03,Bower+07}).  Such a low threshold will naturally
increase the predicted number of transient detections. However, we do
not consider this a realistic threshold since even in the idealized
case of purely Gaussian noise in wide-field radio maps, this will lead
to an unwieldy $\sim 10^3-10^5$ false detections per epoch with ASKAP,
VLA, and SKA; non-Gaussian effects (e.g., \citealt{Frail+12}) may
increase the number of false detections significantly. Moreover, a
$5\sigma$ threshold reduces the dynamic range for detecting the actual
appearance and/or disappearance of a transient, and hence the ability
to separate transients from variable and steady sources.  Finally, the
reduced dynamic range and low signal-to-noise ratio will also inhibit
the ability to infer the physical properties and classification of the
transients (explosion time, energy, ambient density, velocity,
collimation).  Thus, we caution that even if the goal of a survey is
to simply count the number of radio transients (e.g.,
Figure~\ref{fig:lnls}) the detection threshold should be $\sim
10\sigma$.

In principle a transient that does not reach a $10\sigma$ threshold in
any single epoch could be detected at somewhat lower significance in
multiple epochs and hence be recovered.  We essentially account for
this effect through our investigation of the ``stacked'' surveys
(VAST-Wide-Stack, VAST-Deep-SF-STACK, CMB-Stack-1, CMB-Stack-2) in
which we carry out the transient detection in temporally-binned
versions of the nominal daily cadence data.  On the other hand, for
the various surveys with a yearly cadence (e.g., VAST-Deep, VLASS),
which is comparable to or longer than the typical transient durations
(\S\ref{sec:general}), such time-averaging is not feasible and
$10\sigma$ remains a robust single-epoch threshold to suppress a large
number of false detections.

We emphasize that we do not explicitly model the complex processes of
imaging interferometric data, cataloging sources, and measuring
fluxes, which are subject to a variety of systematics that have proven
challenging to cope with (e.g., radio-frequency interference, spurious
sidelobe sources, correlator bugs).  Instead we assume that these
effects can be approximated as a degradation of a survey's limiting
flux density and/or effective area.  Accounting for these systematic
effects is part of our motivation for choosing $10\sigma$ as a robust
detection threshold.  We also assume that each survey achieves its
characteristic sensitivity uniformly on the sky, while in practice the
sensitivity is spatially variable, or, equivalently, the effective
area is a function of sensitivity (e.g.,
\citealt{Croft+13,Williams+13}).

The transient detection criteria in radio surveys should also include
the effects of contamination by radio emission from star-forming host
galaxies. This has the potential to be more of a problem for radio
surveys as compared to optical ones because of the challenges in
subtracting sources with complex substructure in interferometric data
with sparse $uv$ coverage. Transient searches will likely be performed
using flux measurements that include a contribution from the host
galaxy, which will reduce the effective change in brightness of a
transient.

Host galaxy contamination is of greatest concern for sources that
occur exclusively in star-forming galaxies (LGRBs, LLGRBs, Type Ib/c
SNe), but it will also affect some fraction of transients that occur
in all galaxy types (NS-NS mergers, SGRBs, TDEs).  Galaxy
contamination is also a greater problem for lower luminosity
transients (e.g., LLGRBs, Type Ib/c SNe) than for those that easily
outshine their host galaxies (e.g., on-axis LGRBs).  In
Figure~\ref{fig:LCs}, we mark the host galaxy flux density
corresponding to a star formation rate of ${\rm SFR}=1$ M$_{\odot}$
yr$^{-1}$:
\begin{equation}
  F_{\rm \nu,gal} \approx 0.3\,{\rm mJy}\,\,\nu_{\rm GHz}^{-0.7} D_{L,27}^{-2}
  \left(\frac{\rm SFR} {M_{\odot}\,\rm yr^{-1}}\right),
\label{eq:Fgal}
\end{equation}
where we use the mean normalization measured for local galaxies
\citep{Carilli&Yun99} and adopt a spectral index of $-0.7$ relevant
for galactic synchrotron emission which dominates in this frequency
range\footnotemark\footnotetext{The 150 MHz flux derived this way may
  be over-estimated by a factor of $\sim 2$ due to the presence of
  free-free absorption (e.g., \citealt{Williams&Bower10}).}
\citep{Condon92}.  From Figure~\ref{fig:LCs} it is clear that galaxy
contamination will affect a large fraction of transient classes at 150
MHz, but that only LLGRBs, Type Ib/c SNe, NSMs, and off-axis SGRBs
will be affected at $1$--$150$ GHz.

The host galaxy contamination will be mitigated for sources that are
sufficiently nearby such that their hosts will be spatially resolved
by the survey interferometer. However, with the exception of the VLASS
and the SKA 1.3 GHz surveys, the other surveys have large beam sizes
of $\sim 10''$ (VAST and SKA-Low), $\sim 1'$ (LOFAR), and $\sim 14'$
(CMB).  For a typical galaxy size of a few kpc this means that the
hosts will only be significantly resolved for VAST and SKA-Low within
only a small distance of $\lesssim 50$ Mpc, while for LOFAR and the
CMB surveys the hosts will generally be unresolved. Taking into
account the transient luminosities, the survey angular resolutions,
the expected host galaxy brightness at the various frequencies, and
the occurrence rate of transients in star-forming galaxies we indicate
in Tables~\ref{tab:LGRB}--\ref{tab:TDE} whether host galaxy
contamination is expected to be a problem for transient detection.

\subsection{Monte Carlo Method}
\label{sec:mc}

Our method for calculating the detection rates and measured properties
of the various transients with a given survey is as follows.  We
create a large number ($\gtrsim 10^4$) of mock transient light curves
using the models in \S\ref{sec:classes} (Figure~\ref{fig:LCs}), with
random time phasing relative to the start of the survey and with
random distances, weighted by the comoving cosmological volume and,
when appropriate, star formation rate or SMBH density as described in
\S\ref{sec:classes}.  The transient light curves are calculated at
each redshift, taking into account K-corrections and time dilation
effects when necessary.  The detection significance at each epoch is
determined for each event and survey combination, and the detection
criteria are applied to determine the sample of detected
events. Absolute detection rates are finally determined by weighting
the mock sample by the volumetric event rate and the survey duration.
A new realization of events is considered for each survey separately.
For each survey we also quantify the fraction of events ($f_V$) that
pass the flux threshold criterion but are miseed by the factor of two
variability cut.  This fraction provides a measure of how many
transients could in principle be recovered with a much longer survey
duration.

We note that the issue of detection alone is separate from the question of
what qualities make a given event scientifically useful. We defer a detailed
discussion of this issue to Paper II. However, some insight into this question
can be gained based on a few properties that are readily available from the
mock transient sample. In particular, for the transient sources that pass the
detection criteria, we determine the number of epochs with $\gtrsim 3\sigma$
detections to assess the number of light curve data points. We also determine
whether the source is detected to rise and/or decline by at least a factor of
two relative to the measured light curve peak. If a source is only detected to
rise or to decline during the survey, then the only possible measurement is a
lower limit on the peak flux and on the rise or decline time, respectively. If
the transient is detected at greater than half of its maximum flux at either
the first or last epoch of the survey, then the rise or decline time cannot be
determined, respectively. These operational definitions are important because
unlike in the case of radio follow-up of transients discovered at other
wavelengths (e.g., GRBs, SNe, TDEs), the initial time of the event is not
known a priori. In \autoref{fig:zdistrib} we present the cumulative redshift
distributions of the transients discovered by the SKA surveys.

\section{Simulation Results}
\label{sec:results}

\begin{figure*} \includegraphics[width=0.5\textwidth] {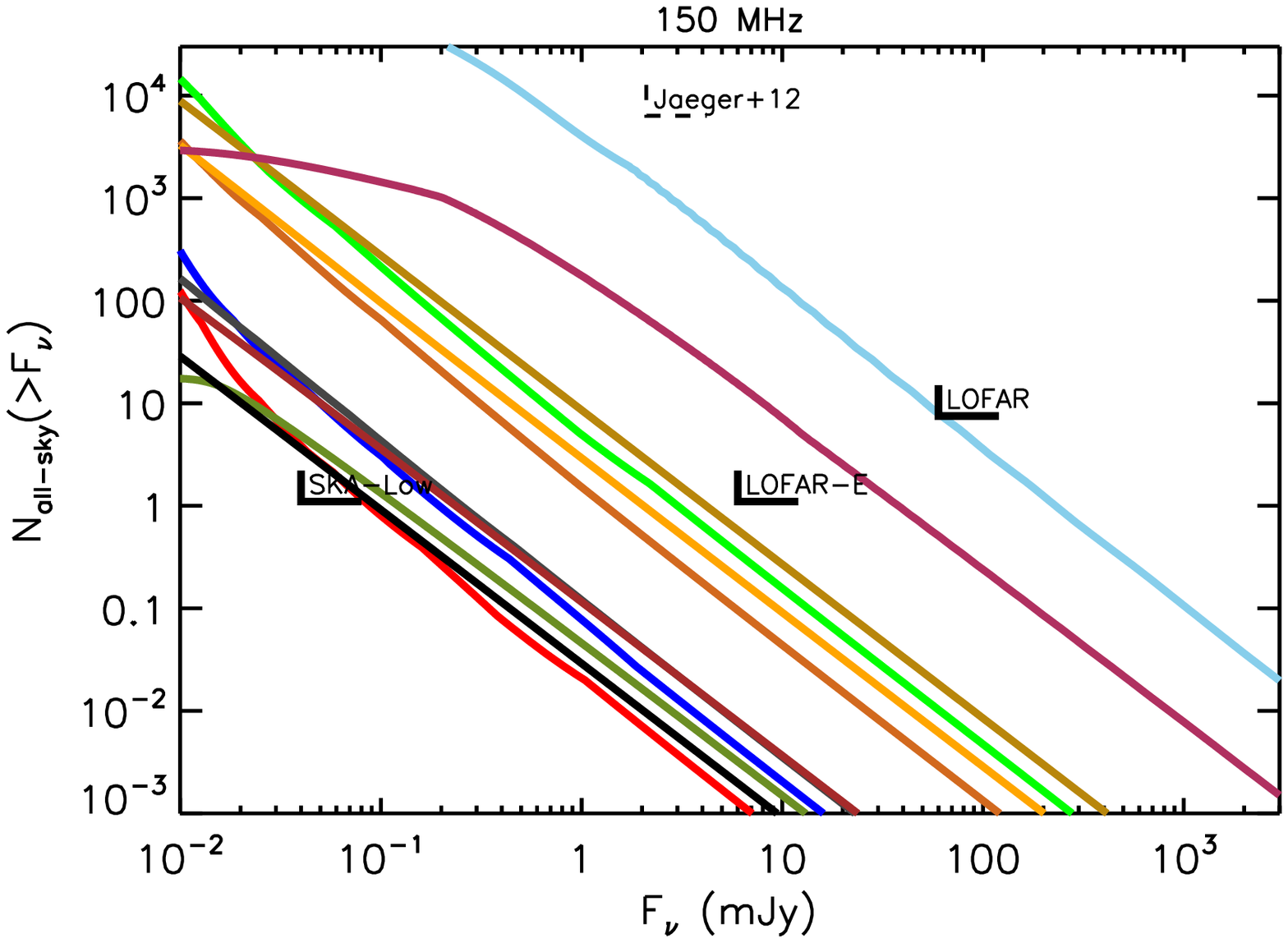}
  \includegraphics[width=0.5\textwidth] {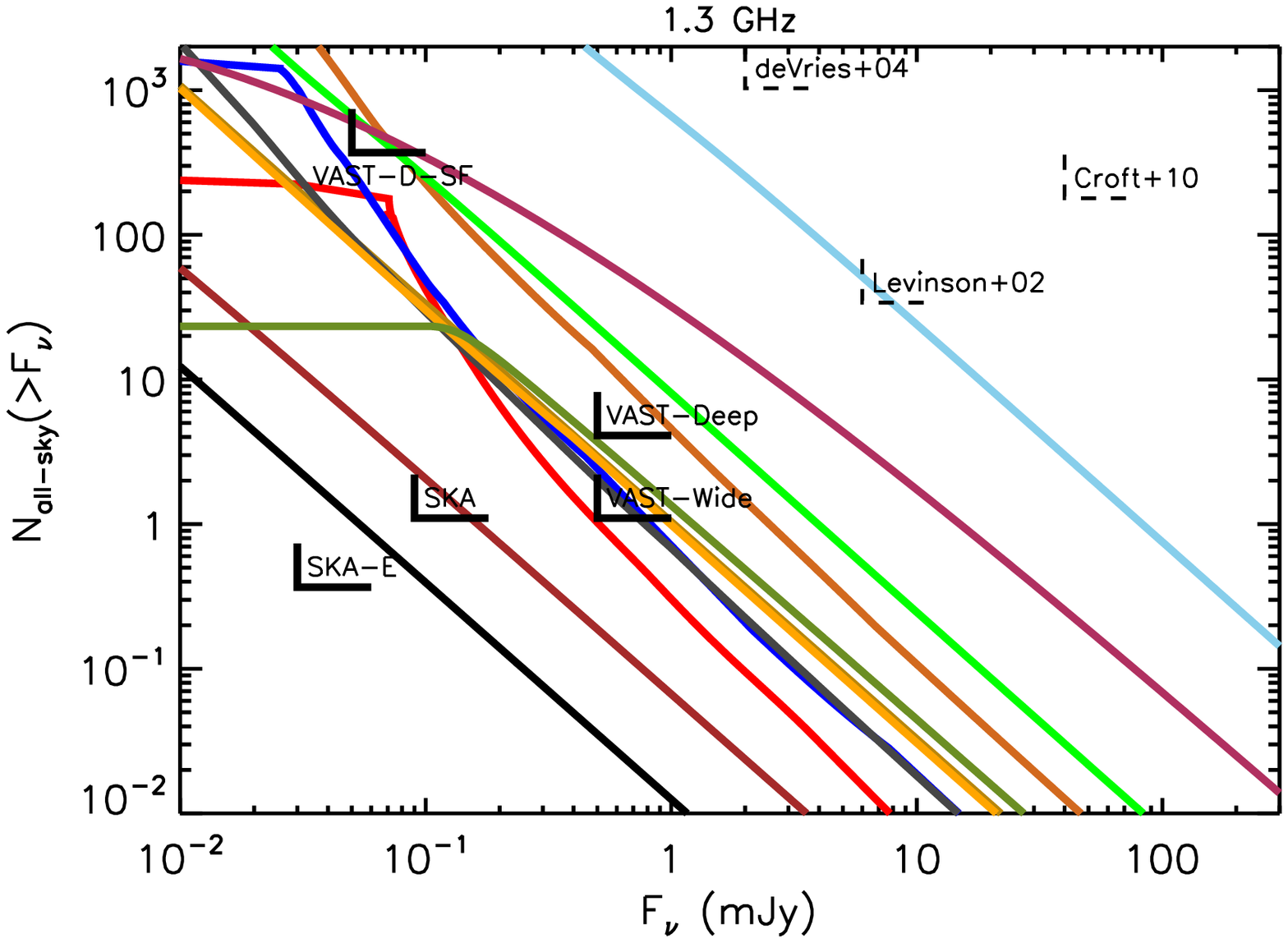}
  \includegraphics[width=0.5\textwidth] {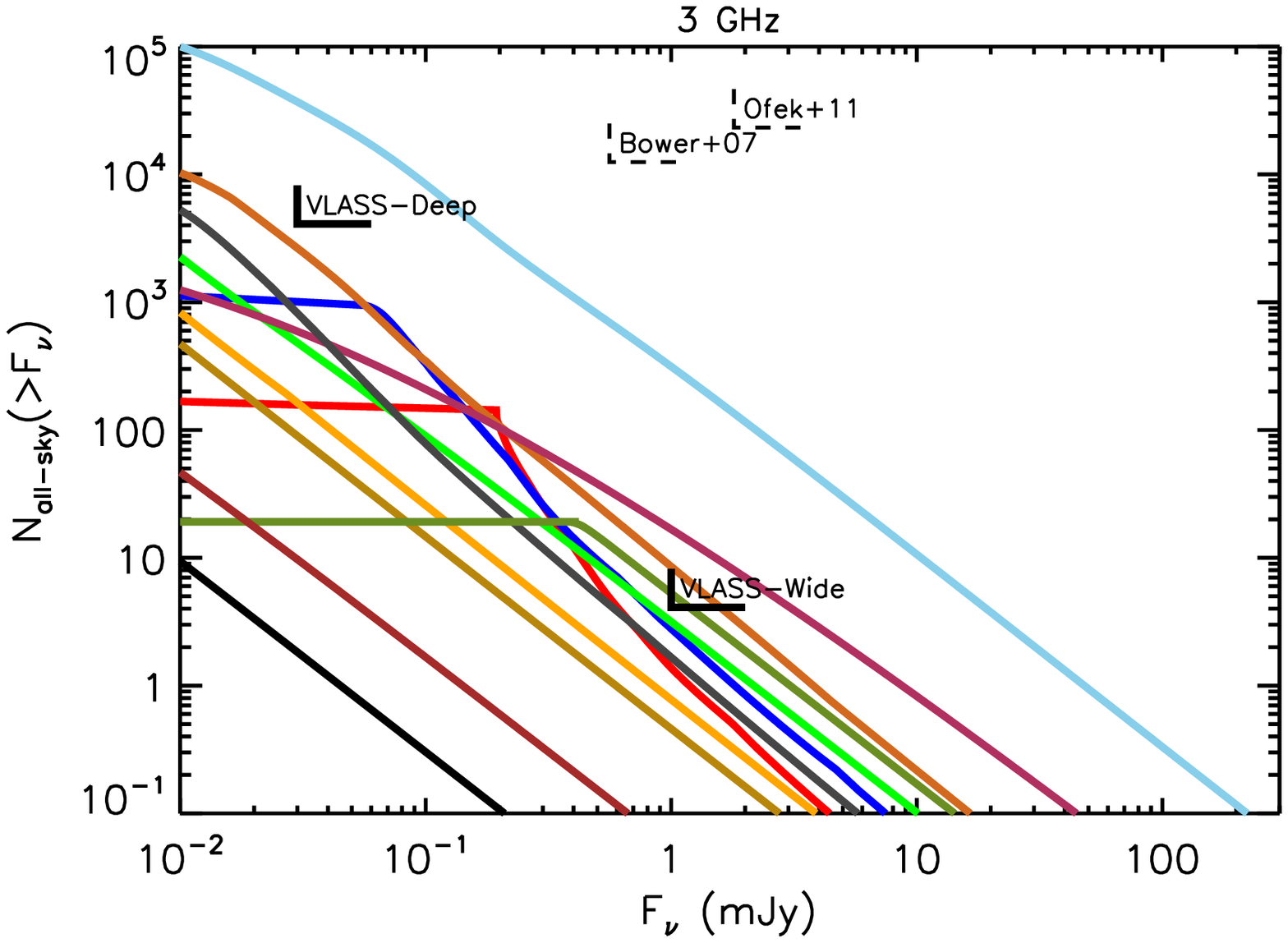}
  \includegraphics[width=0.5\textwidth] {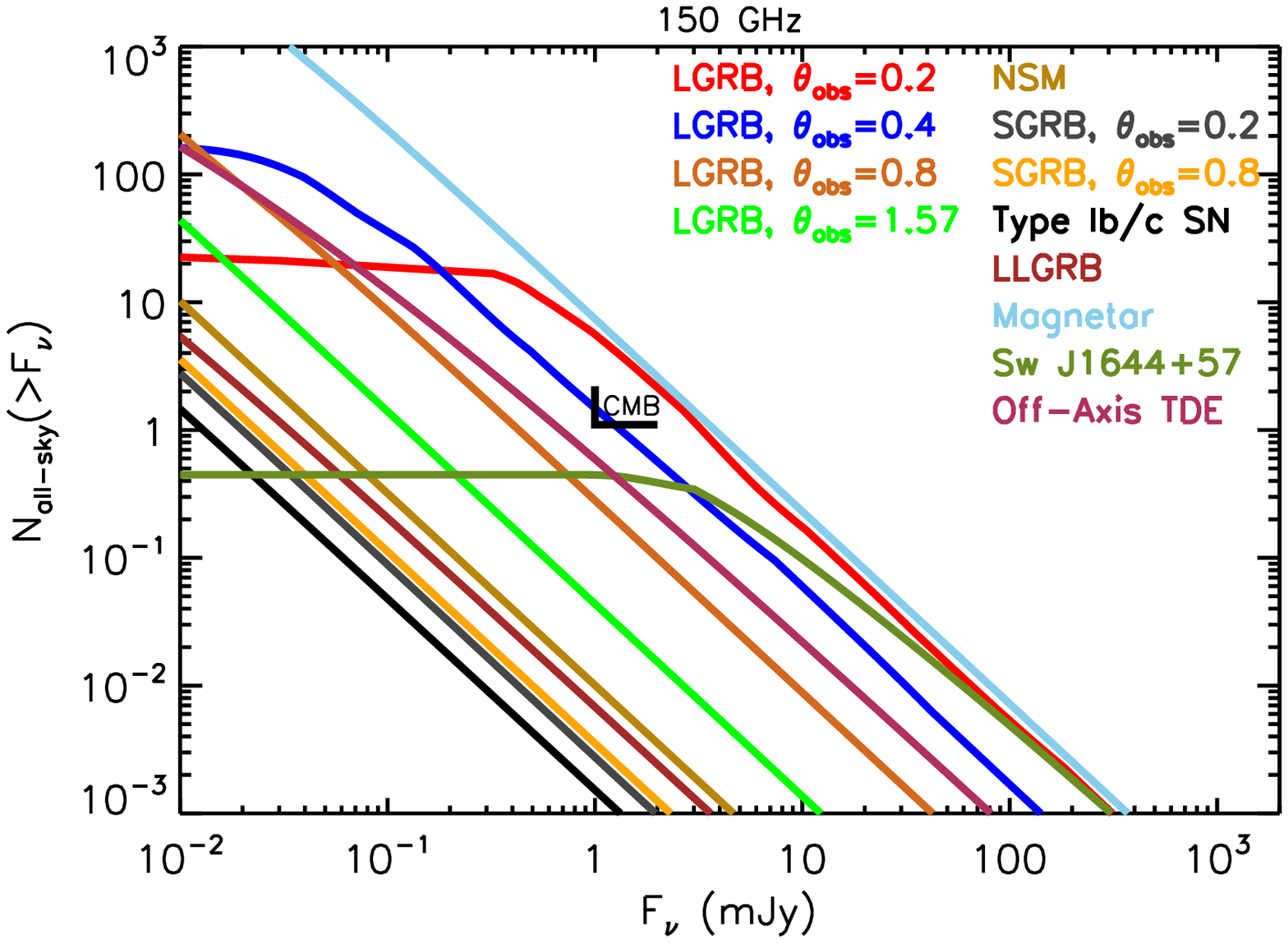}
  \caption{ Number of sources across the entire sky
    (eq.~[\ref{eq:Ns}]) at any time above flux $F_{\nu}$ at observer
    frequency $\nu = 150$ MHz (upper left), 1.3 GHz (upper right), 3
    GHz (lower right), 150 GHz (lower right). Sources shown include
    long GRBs ($\theta_{\rm obs}$ = 0.2, {\it red}; $\theta_{\rm obs}$
    = 0.4, {\it blue}; $\theta_{\rm obs}$ = 0.8, {\it brown};
    $\theta_{\rm obs}$ = 1.57, {\it green}), neutron star merger
    leaving black hole ({\it olive}), NSM-magnetar ({\it light blue}),
    on-axis SGRB ({\it grey}), off-axis SGRB ({\it orange}), on-axis
    TDE (\swsixteen; {\it dark green}), off-axis TDE ({\it purple}).
    Shown for comparison are representative limits placed by the
    surveys discussed in this paper (Table \ref{tab:surveys}) as well
    as those placed by past surveys (Appendix \ref{sec:limits};
    references: \citealt{Jaeger+12},\citealt{deVries+04},
    \citealt{Croft+10}, \citealt{Levinson+02}, \citealt{Gal-Yam+06},
    \citealt{Ofek+10}, \citealt{Ofek+11}, \citealt{Bower+07}; see also
    \citealt{Frail+12}).}
\label{fig:lnls}
\end{figure*}

\begin{figure*} \includegraphics[width=\textwidth] {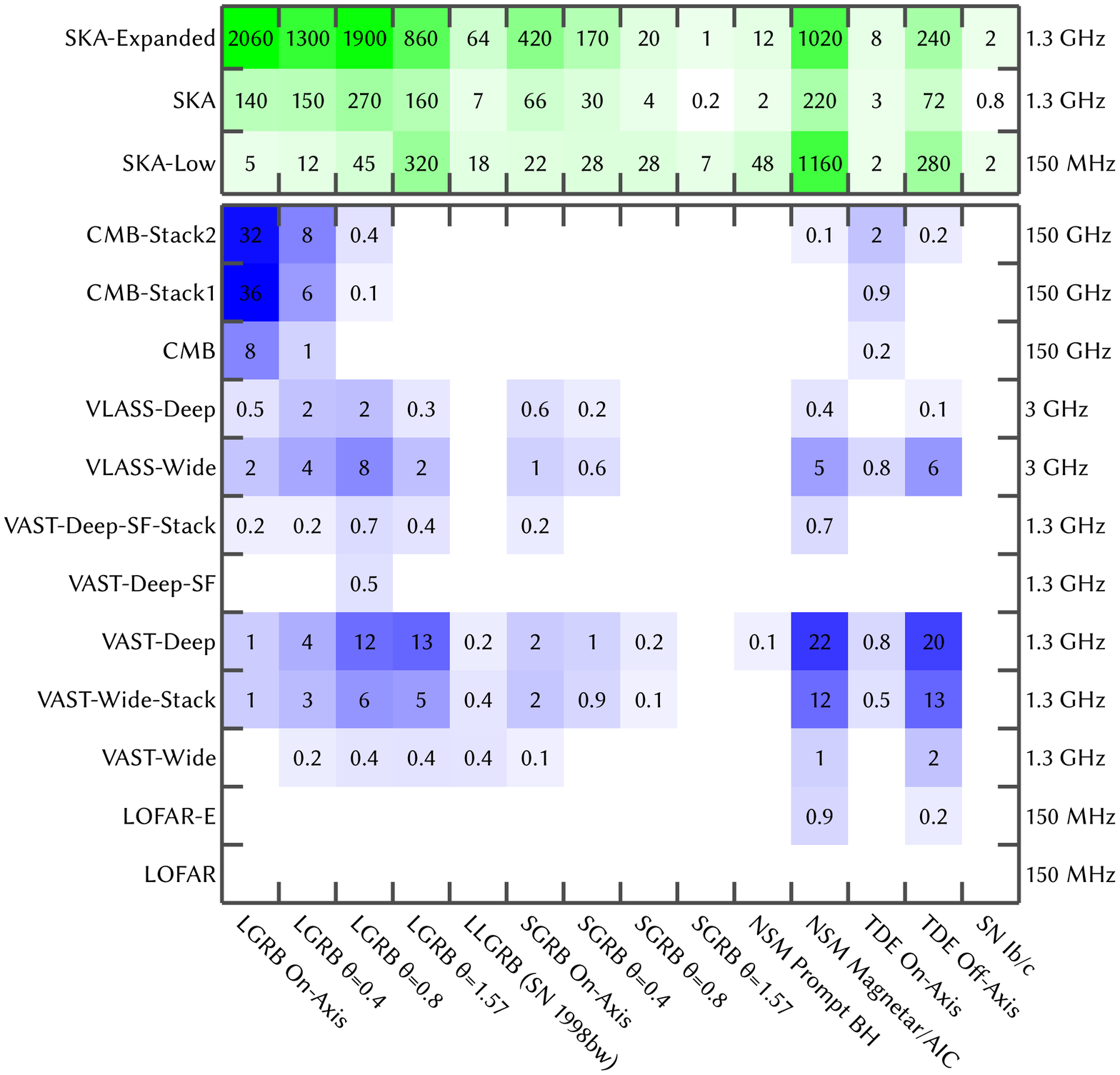}
  \caption{Predicted number of event detections ($N$ in
    Tables~\ref{tab:LGRB}--\ref{tab:TDE}) by survey and event class. Predicted
    values less than 0.1 are not shown. Because of the varying quality of
    constraints on the underlying rate and light curve models, uncertainties
    in $N$ vary from one event class to another (\S\ref{sec:models}).}
\label{fig:yields}
\end{figure*}

\begin{figure*}
  \includegraphics[width=0.5\textwidth] {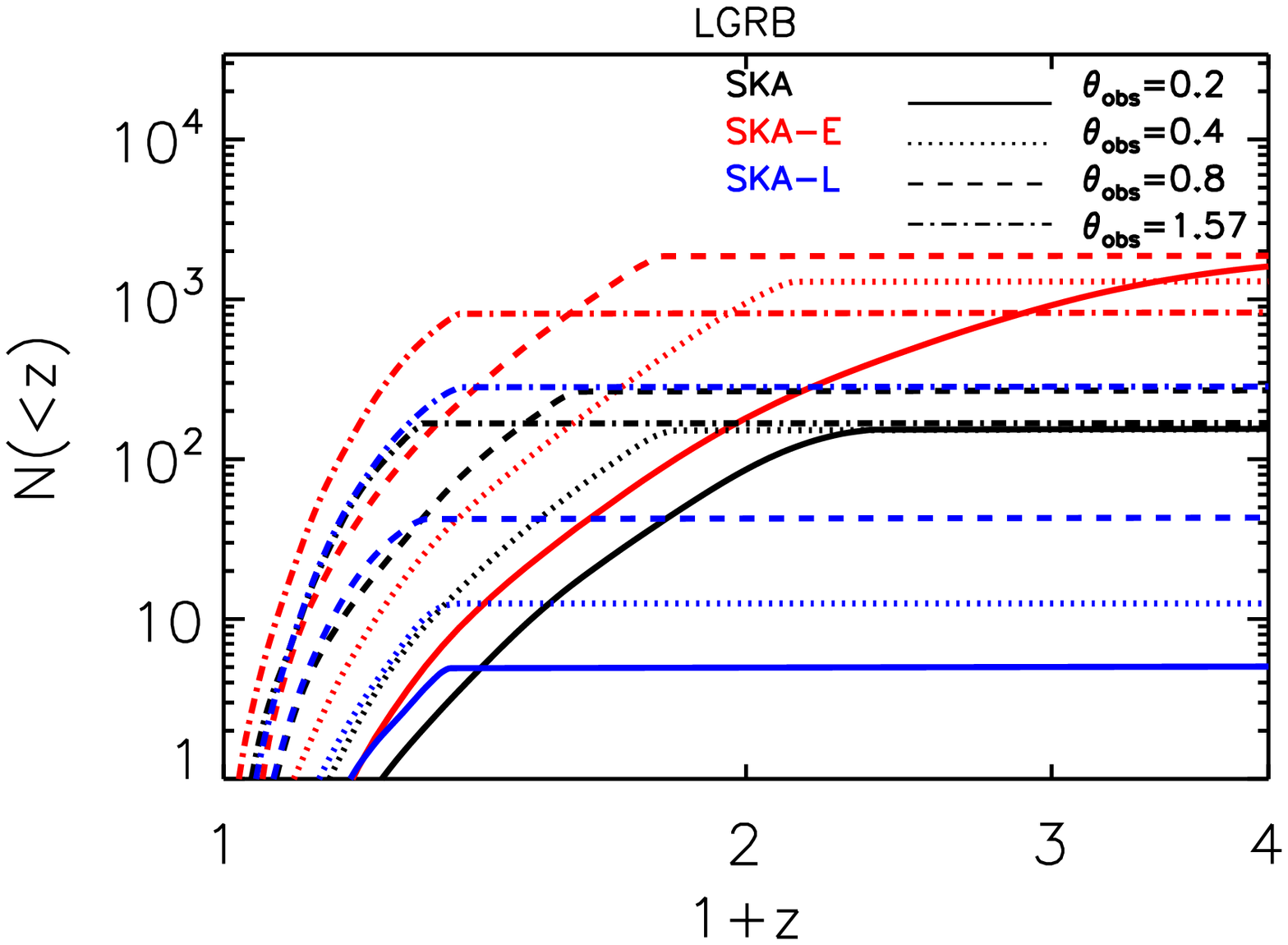}
  \includegraphics[width=0.5\textwidth] {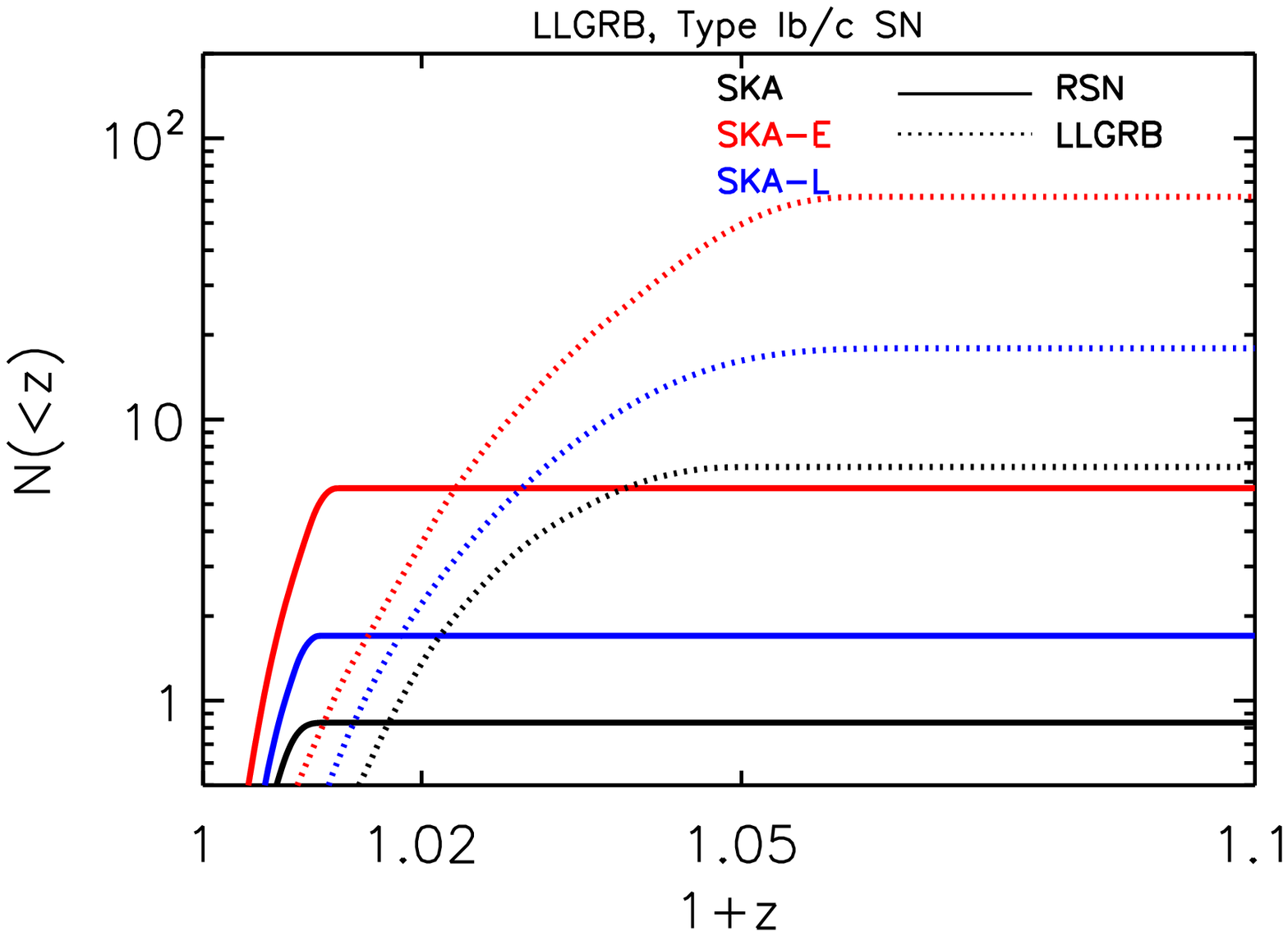}
  \vskip 10pt
  \includegraphics[width=0.5\textwidth] {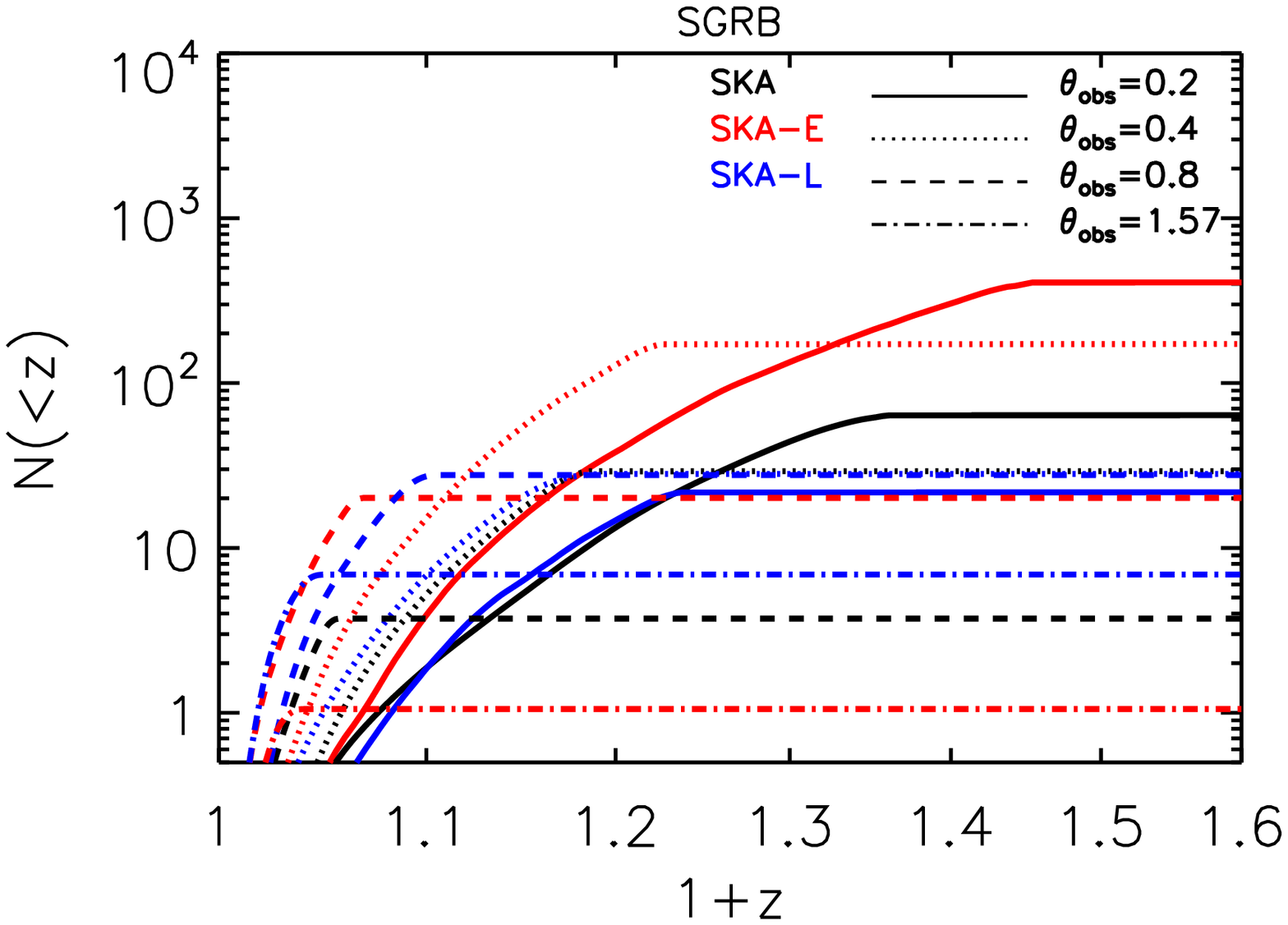}
  \includegraphics[width=0.5\textwidth] {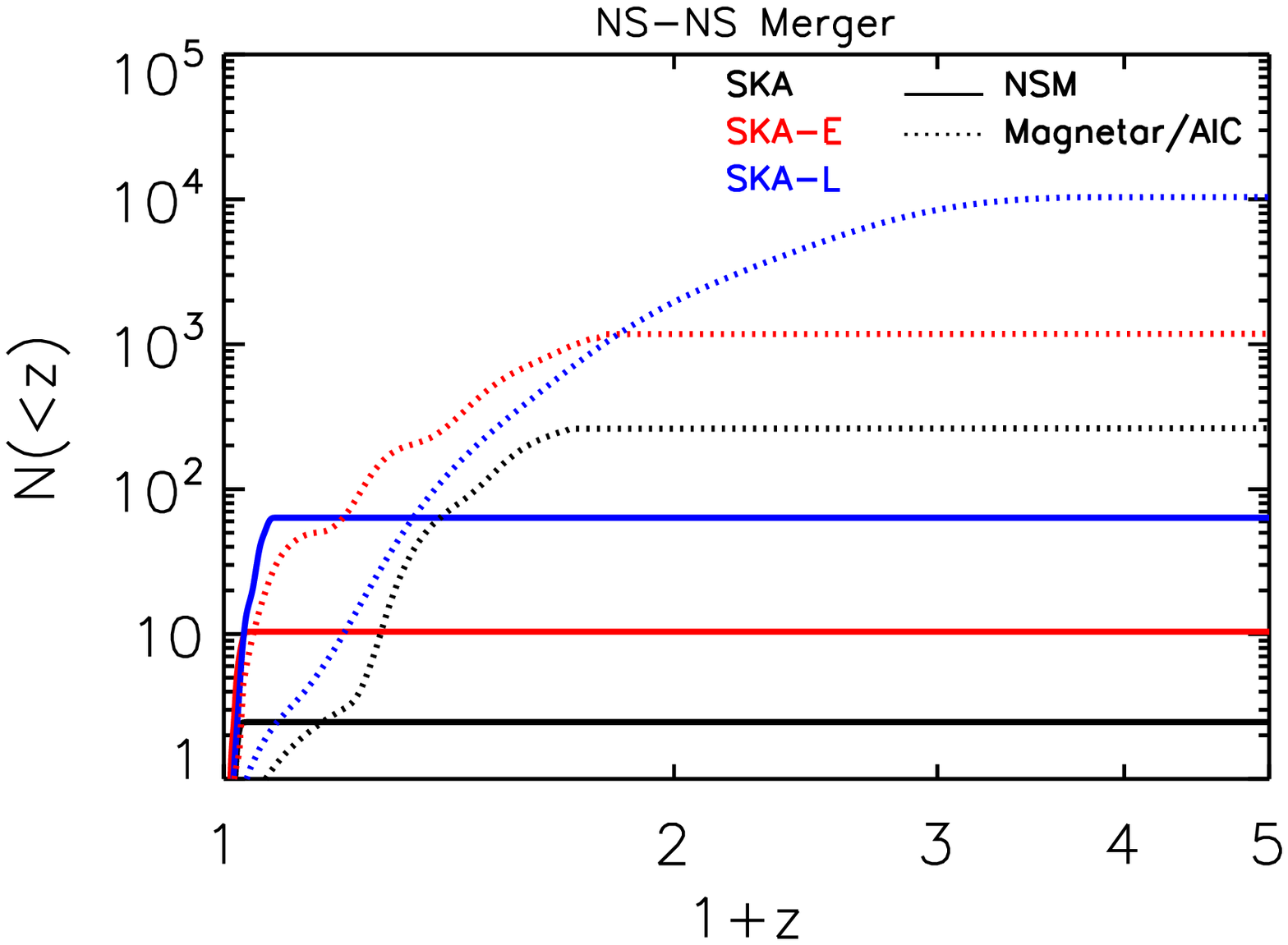}
  \vskip 10pt
  \includegraphics[width=0.5\textwidth] {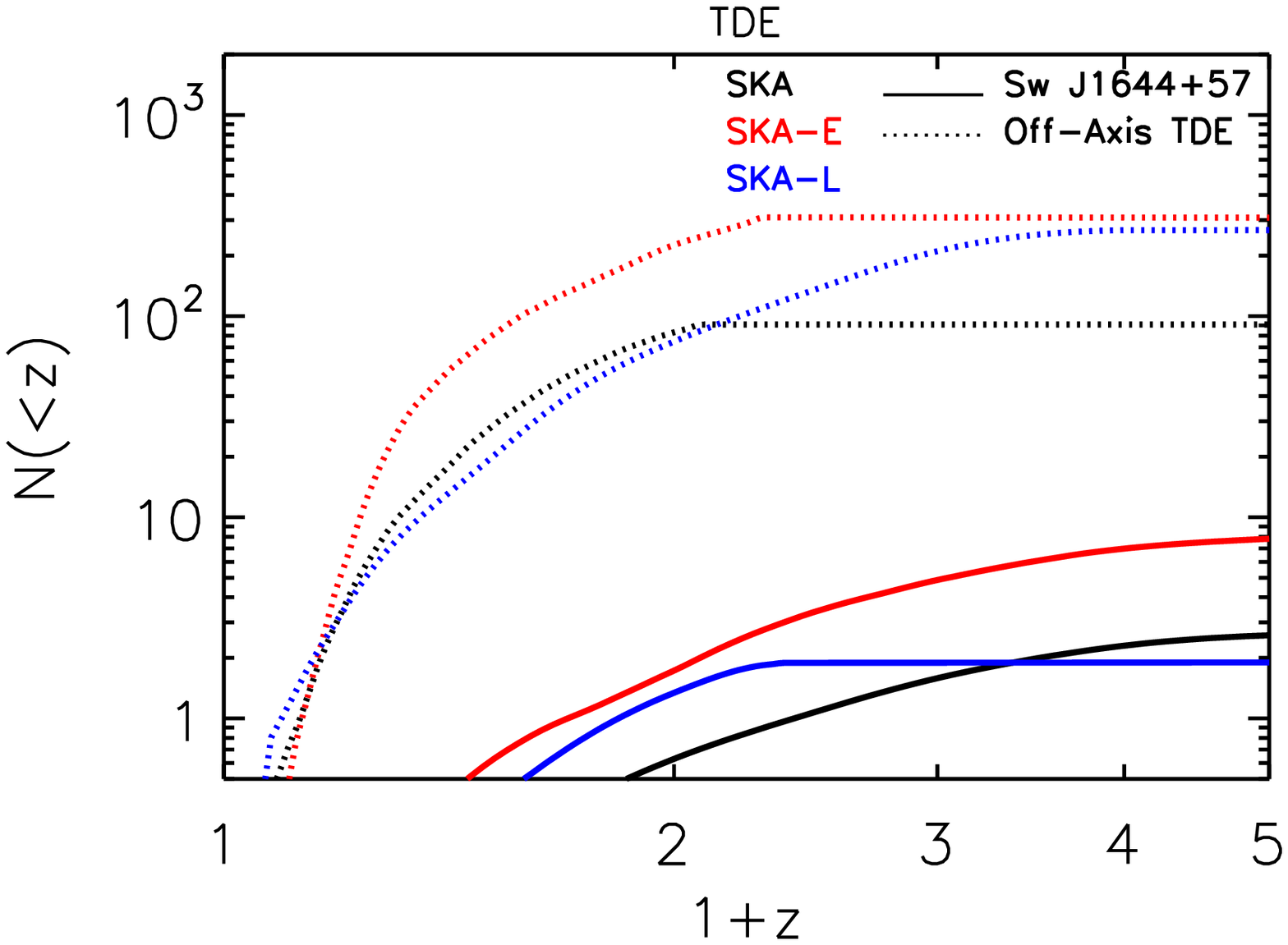}
  \caption{Cumulative number of detected sources with redshift less than $z$
    for each of the three SKA surveys considered in this paper (SKA: {\it
      black}, SKA-Expanded: {\it red}, SKA-Low: {\it blue}). Different source
    classes are shown in different panels and with different line styles as
    shown in the legends. Note that the top right panel shows results for both
    LLGRBs and SNe Ibc.}
\label{fig:zdistrib}
\end{figure*}

The results of our Monte Carlo simulations are summarized in
Tables~\ref{tab:LGRB}--\ref{tab:TDE} and
Figures~\ref{fig:lnls}--\ref{fig:zdistrib}.  The Tables provide the
number of detected transients for each class and survey ($N$), the
mean number of epochs at which a detected transient has a flux density
of $\gtrsim 3\sigma$ ($\bar{n}$), the mean redshift of the detected
sources ($\bar{z}$), the fractions $f_{\rm rise}$ and $f_{\rm fall}$
of detected transients with ``measured'' rise and decline times (as
defined in \S\ref{sec:mc}), respectively, and the fraction of detected
transients with their peak occurring during the survey, $f_{\rm peak}$.
For each transient class and survey we also estimate whether galaxy
contamination will be an issue (`Yes', `No', or `Maybe').  These
designations are based on the transient peak brightness relative to
the host galaxy radio emission (for a fiducial ${\rm SFR}=1$ M$_\odot$
yr$^{-1}$), as well as on whether all events in a given class are
expected to occur in star-forming galaxies.  Figure~\ref{fig:yields}
summarizes the total number of detections for each transient class by
each survey.  

Before addressing the individual results in detail, we consider in
Figure~\ref{fig:lnls} the number of sources detected across the entire
sky at a fixed point in time as a function of the survey depth
($10\sigma$) for each radio frequency and transient class.  The number
of sources above a given flux density is estimated according to:
\begin{equation}
  N_{\rm all-sky}(>F_{\nu}) = \int_{0}^{z(F_{\nu,\rm p})}
  \frac{\mathcal{R}(z')t_{\rm dur}(z')}{1+z'} dV',
\label{eq:Ns}
\end{equation} 
where $\mathcal{R}(z')$ is the co-moving volumetric density, $t_{\rm
  dur}(z')$ is the transient duration at redshift $z$ (estimated as
the timescale over which the flux is greater than one half of its peak
value), the $(1+z)^{-1}$ factor accounts for time dilation of the
event rate, and the integral is performed over the co-moving volume
$V$ out to the redshift $z(F_{\nu})$ corresponding to the peak flux
density $F_{\nu,\rm p}$.  We stress that since this is a snapshot
rate no variability cuts are applied, and therefore there is no
guarantee that all of these sources can actually be identified as
transients in a real survey\footnotemark\footnotetext{For instance, at
  1.3 GHz, $N_{\rm all-sky}(>F_{\nu})$ is at least an order of
  magnitude larger for the NSM-magnetar case than all other events,
  but we do not predict a commensurate number of discoveries in actual
  surveys, because $\sim 75\% $ of these events do not vary
  substantially over the survey duration
  (\autoref{tab:NSM-magnetar}).}.

As can be seen in Figure~\ref{fig:lnls} all transient classes follow
the relation $N_{\rm all-sky} \propto F_{\nu}^{-3/2}$ at high flux
densities, consistent with the expectation for Euclidean geometry and
a homogeneous source population.  However, significant deviations are
seen at low flux densities due to a combination of K-corrections,
non-Euclidean luminosity distance, and cosmological evolution of the
source population.  These effects have not been systematically
considered in previous studies.  For luminous sources, $N_{\rm
  all-sky}(>F_\nu)$ is seen to flatten below a critical flux density,
corresponding to the depth at which essentially all events are visible
to the edge of the Universe; for such sources deeper searches are
ineffective at increasing the detection rate.  Whether the slope of
the distribution flattens or steepens approaching this plateau depends
on the importance and sign of the K-correction.  An upward steepening
of $N_{\rm all-sky}(>F_\nu)$ occurs for sources with a strong negative
K-correction, which is due to the steep positive spectrum of
synchrotron self-absorption ($F_{\nu} \propto \nu^2$ or $\propto
\nu^{5/2}$); e.g., on-axis LGRBs at $\lesssim 3$ GHz.  On the other
hand, a flattening prior to the plateau occurs when the spectrum is
flat or inverted; e.g., on-axis LGRBs at 150 GHz.

Shown for comparison in Figure~\ref{fig:lnls} are available
constraints from previous radio surveys, as well as from the surveys
we simulate here.  The detection limits are calculated at the $95\%$
confidence level for a fiducial transient duration of $\sim 100$ days
(as motivated in \S\ref{sec:general}).  The methodology for
calculating the upper limits, and a summary of the results from past
surveys are given in Appendix~\ref{sec:limits}.  With the possible
exception of NSM-magnetar events (\S\ref{sec:magnetar}), which at our
fiducial rates approach the constraint at 1.4 GHz set by FIRST/NVSS
\citep{Levinson+02}, none of the previous surveys reach the required
depth/sky area to detect any of the known classes of extragalactic
transients, consistent with the lack of confirmed transients to date
\citep{Frail+12}.

\subsection{LGRBs}

The VAST (Deep, Wide, and Wide-Stack), VLASS, CMB, and SKA surveys all
detect at least one LGRB. 

The wide/shallow surveys VAST-Deep, VAST-Wide-Stack, and VLASS-Wide
detect $\sim 1-2$ on-axis LGRB afterglows, but this number increases
to $\sim 15-30$ for off-axis events.  The off-axis events are
generally detected over a few epochs, with a low mean redshifts,
$\bar{z}\sim 0.1$.  In addition, $\sim 50-80\%$ will have measured
rise and decline times.  These GHz surveys should thus produce a
reasonably complete sample of LGRB afterglows which are located
sufficiently nearby to possess detectable SNe (detected archivally
since the SN optical emission will probably have faded by the time of
the first radio detection).  However, we note that the SNe may be
identified first in the optical, in which case targeted radio
follow-up of such events presents a more profitable approach than an
untargeted wide-field radio search.

The wide/shallow surveys detect a greater number of LGRBs than the
deeper, pencil-beam surveys VLASS-Deep and VAST-Deep-SF-Stack, which
lead to $N\sim 1-4$ off-axis events. However, the latter do better
than predicted from a homogeneous source population because the LGRB
volumetric rate increases by an order of magnitude at the larger
redshifts ($z\gtrsim 1$) to which these surveys are sensitive.

The stacked CMB surveys detect $N\approx 40$ events with an average redshift
of $\bar{z}\approx 0.7$. The long effective cadence ($\sim 100$ days) of the
CMB-Stack-2 survey compared to the typical LGRB afterglow duration at 150 GHz
renders most events detectable at only 1 epoch, but each detection by the
CMB-Stack-1 survey ($\sim 10$ day cadence) will occur over $\sim 3$ epochs. In
the latter case, the explosion time of the GRB will typically be constrained
to $\lesssim 10$ days, sufficiently early to trigger optical follow-up and
lower-frequency radio observations to confirm the afterglow origin and to
better localize the event. LGRBs are sufficiently rare and bright at
$\bar{z}\approx 0.7$, that the detection or non-detection of $\gamma$-rays
from the sky location and explosion time window could constrain the existence
of on-axis afterglows that are unaccompanied by a GRB (so-called ``dirty
fireballs'') given the bias towards detecting on-axis events at high observing
frequencies.

The number of on-axis LGRBs detected by millimeter surveys may be
underestimated since we have only modeled emission from the forward shock,
neglecting contributions from the reverse shock which are important at early
times. At 150 GHz the reverse shock emission could be a factor of $\gtrsim 10$
times brighter than the forward shock, but lasting for only $\sim 1$ day
(e.g.~\citealt{Laskar+13}). Approximately 8 on-axis TDEs were detected by the
1-day cadence CMB survey to redshift $\bar{z} \sim 0.34$ considering only the
forward shock emission. From Figure \ref{fig:LCs}, the number of events in the
survey with detectable reverse shock emission could thus be a factor $\sim 20$
times larger, and extending to much higher redshifts $\bar{z} \gtrsim 2$.
Alternatively, LGRBs that are detected by the stacked CMB surveys via their
forward shock emission could be searched at early survey epochs for reverse
shock emission, providing tighter constraints on the time of the GRB and a
probe of the magnetization of the ejecta (e.g.~\citealt{Giannios+08}).

The large SKA and SKA-Expanded surveys at $1.3$ GHz detect $\sim 700$ and
$\sim 6000$ LGRBs, with about $1/3$ of the events viewed on-axis. On-axis
events represent a larger fraction as compared to VAST/VLASS since they are
detected to a higher average redshift, $\bar{z}\sim 1-2$, where the LGRB
volumetric rate is an order of magnitude larger than locally, while the
off-axis events are still mainly nearby ($\bar{z}\sim 0.5$;
\autoref{fig:zdistrib}). SKA-Expanded can detect on-axis LGRBs out to $z > 3$.
The light curves of most detected sources are reasonably well-sampled, with
$\bar{n}\sim 3-27$ epochs, depending on the survey and viewing angle. SKA-Low
detects $\sim 400$ LGRBs, mostly far off-axis ($\theta_{\rm obs}\sim \pi/2$)
and originating from $\bar{z}\lesssim 0.3$.

We note that our predicted off-axis LGRB detection rates are
approximately a factor of $\sim 3$ times smaller than those predicted
by \citet{Ghirlanda+14}, who find that ASKAP VAST should have a
detection rate of $\sim 3\times 10^{-3}$ deg$^{-2}$ yr$^{-1}$,
corresponding to $\sim 90$ events for $10^4$ deg$^{2}$ and a 3 year
survey duration. This discrepancy results largely from our difference
in detection threshold: \citet{Ghirlanda+14} adopt a $5\sigma$
detection threshold, which leads to a factor of $\sim 3$ times higher
predicted rate.

\subsection{LLGRBs}

The SKA surveys are the only ones expected to detect LLGRBs
(Table~\ref{tab:LLGRB} and Figure~\ref{fig:yields}). This is due to the low
luminosity of these events ($\sim 10^2-10^3$ times dimmer than off-axis LGRB
afterglows), which cannot be overcome by their order of magnitude higher
volumetric rate. SKA-Low and SKA-Expanded detect $\sim 20$ and $\sim 60$
LLGRBs at $D_L\lesssim 150$ Mpc (\autoref{fig:zdistrib}). However, host galaxy
contamination could be crippling in the case of SKA-Low due to its relatively
poor angular resolution of $11''$, coupled with the high galaxy flux density
at 150 MHz. We further note that at the mean distance of 150 Mpc, the
broad-lined Type Ib/c SNe, which thus far appear to accompany all LLGRBs, will
be discovered independently much earlier in the optical. On the other hand,
radio-discovered LLGRBs without bright accompanying SNe will be of great
interest.

\subsection{SGRBs}

At most $2-3$ SGRBs are detected by the wide/shallow surveys including
VAST-Wide-Stack, VAST-Deep, and VLASS-Wide, with a high probability of
being close to on-axis ($\theta_{\rm obs}\lesssim 0.4$).  No events
are detected at larger off-axis angles. Much larger numbers, $\sim
100,\,\, 90,\,\, 600$ are detected by the SKA, SKA-Low, and
SKA-Expanded, respectively.  These events are detected in a modest
number of epochs ($n\sim 3-10$), providing reasonably well-constrained
rise and decline times.

The mean redshift of the SGRBs detected by the SKA surveys, $\bar{z}
\sim 0.2$, is sufficiently high that most events will not have
associated gravitational wave detections with Advanced LIGO/Virgo (see
also \citealt{Metzger&Berger12}, who reach a similar conclusion based
directly on the local SGRB rate).  For the few events within the
gravitational wave detection range, the (albeit poor) sky
localizations from Advanced LIGO/Virgo will be sufficient to uniquely
associate the gravitational wave source with the SGRB radio afterglow,
even given a $\sim {\rm year}$ uncertainty in coincidence time.

\subsection{NS-NS Mergers with Prompt BH Formation}

VLASS-Wide-Stack, VAST-Deep, and VLASS-Wide detect $\lesssim 0.1$
NS-NS mergers, although this number is uncertain by at least an order
of magnitude given the uncertain local merger rate.  These numbers are
suppressed by an order of magnitude because a substantial fraction
($f_{\rm V} \sim 0.8-0.9$) of nominal detections are lost to the
variability cut, due to the slow evolution of NS-NS merger light
curves relative to the survey durations (Figure~\ref{fig:LCs}).  SKA
and SKA-Expanded detects $\sim 2$ and $\sim 10$ events, respectively,
with $\bar{z}\sim 0.02$, while SKA-Low detects $\sim 50$ with $\bar{z}
\lesssim 0.05$.  All detected events are sufficiently close for
Advanced LIGO/Virgo to detect their gravitational wave emission, which
will happen well in advance of the radio light curve maxima.

\subsection{NS-NS Mergers with Stable Magnetar Remnants \& White
  Dwarf AIC}

NSM-magnetar events are among of the most promising transient classes due to
their high luminosities. The wide/shallow surveys VAST-Wide-Stack, VAST-Deep,
and VLASS-Wide detect $\sim 10,\,\, 20, \,\, 5$ events, respectively to an
average redshift of $\bar{z}\sim 0.2-0.4$. Due to the long timescales of these
events, almost all detections occur at every epoch, and the peak is measured
in only $\sim 30-60\%$ of the cases. SKA, SKA-Expanded, and SKA-Low detect
$\sim 200,\,\, 1000,\,\, 1200$ events, respectively, to much higher redshifts
of $\bar{z}\sim 0.8-1.4$. Unlike the shorter-wavelength surveys, SKA-Low has
the capability to detect NSM-magnetar events to $z \gtrsim 2$
(\autoref{fig:zdistrib}). We stress that while the optimistic detection rates
make these sources of great interest for future radio surveys, the fraction of
mergers that lead to stable magnetars (as well as the overall NS-NS merger
rate) is highly uncertain. Upcoming surveys will constrain the true occurrence
rate.

NSM-magnetar represent the kind of source for which untargeted radio transient
surveys have more discovery potential \citep{Metzger&Bower14} because they are
expected to be radio-luminous without having counterparts that are more
readily discovered at other wavelengths. These events may also be accompanied
by luminous X-ray and optical emission following the merger or AIC (e.g.,
\citealt{Metzger+08,Yu+13,Metzger&Piro14}), but the short duration and low
luminosities of these putative emission processes will make their discovery
challenging as compared to at radio frequencies.

\subsection{On-Axis Jetted TDEs (\swsixteen)}

CMB-Stack-2 detects $\sim 2$ on-axis TDE, with $\bar{z}\sim 1$, while
the wide/shallow GHz surveys VAST-Wide-Stack, VAST-Deep and VLASS-Wide
detect $\lesssim 1$ on-axis TDEs (Table~\ref{tab:TDE}).  While the
peak flux density of these events is higher at 3 GHz than at 1.3 GHz,
VLASS loses a greater number of events to the variability cut ($f_{\rm
  V} \sim 0.5$) as compared to VAST ($f_{\rm V} \sim 0.3-0.4$).  The
pencil beam survey VLASS-Deep detects fewer events than would be
predicted even assuming a Euclidean geometry and homogeneous source
population because on-axis TDEs are so luminous that their detection
rate is limited not by the flux density threshold of the survey, but
by the decreasing SMBH density at $z\gtrsim 4$ and the greater loss of
events due to the variability cut introduced by time dilation ($f_{\rm
  V} = 0.75$).  The same effect limits the number of detections by the
various SKA surveys to $\lesssim {\rm few}$, despite their much
greater sensitivity.  In essence, on-axis jetted TDEs are rare events,
even if one can probe the entire observable universe.

The long durations of jetted TDE afterglows result in most events
being detected in most survey epochs, e.g., $\bar{n}\approx 7$ for
CMB-Stack-2, yet the duration is sufficiently short that most will
have measured rise times.  Although the initial epoch of jet launching
will only be constrained by radio observations to within $\sim 100$
days, the rarity of jetted TDEs implies a good chance to unambiguously
associate the radio sources with prompt gamma-ray emission detected by
{\it Swift} or {\it Fermi}, either via a previous trigger or through
an archival search (e.g., \citealt{Cenko+12}).  The soft X-ray
emission might also still be detectable, as luminous X-rays
accompanied \swsixteen\ for about 500 days after the TDE
\citep{Zauderer+13}.  We also note that the radio detection rates
could be higher than our predictions if not all radio-producing
on-axis TDEs are accompanied by luminous X-ray emission, since our
rates are derived from X-ray-discovered events.  However, the
typically high redshift of detected events of $\bar{z} \gtrsim 0.6$
implies that constraining the transient location to the nucleus of the
host galaxy with sufficient accuracy to confirm a TDE origin will be
challenging, likely requiring follow-up JVLA or VLBA imaging.

\subsection{Off-Axis Jetted TDE}
\label{sec:offaxistderesults}

Off-axis jetted TDEs exhibit higher detection rates, with $\approx 13,\,\,
20,\,\, 6$ events detected by VAST-Wide-Stack, VAST-Deep and VLASS-Wide,
respectively (Table~\ref{tab:TDE}). These events occur at lower average
redshift ($\bar{z}\sim 0.2-0.3$; \autoref{fig:zdistrib}) than the on-axis
events and are also detected at most epochs of the survey ($\bar{n}\approx
N_{\rm obs}$). Detected events are reasonably well-characterized, with rise
and decline times determined for $\sim 70\%$ and $\sim 30\%$ of events,
respectively. The typically large viewing angle with respect to the jet axis
and relatively small typical distances imply that VLBI imaging of off-axis
TDEs could allow the extended jet structure to be resolved
\citep{Giannios&Metzger11,Mimica+15}, potentially providing an independent age
for the system.

LOFAR-Extended detects less than one event ($N\approx 0.2$) due in
part to the large number of events lost to the variability cut
($f_{\rm V} \approx 0.9$).  The millimeter surveys also detect
$\lesssim 1$ off-axis TDEs due to the suppressed peak flux at high
frequencies as compared to the on-axis case.

We note that our overall TDE detection rates are much lower than those
predicted by \citet{Frail+12}, primarily because these authors assumed
a beaming correction to the \swsixteen\ rate of $10^3$ (instead of
$10^2$ here).  They also assumed that the off-axis TDE light curve at
5 GHz is identical to the on-axis case (\swsixteen), whereas at 3 GHz
we find that the peak flux density is a factor of 6 times dimmer in
the off-axis case (Figure~\ref{fig:LCs}).  Our assumptions result in a
predicted TDE radio detection rate that is two orders of magnitude
lower than in \citet{Frail+12}.

\subsection{Type Ib/c SNe}

No Type Ib/c supernovae are detected in any of the surveys
(Table~\ref{tab:LLGRB}), except perhaps the SKA surveys, which may
detect a few events.  As in the case of LLGRBs, these low detection
rates result from the low brightness of these events ($\gtrsim 10^3$
times dimmer than off-axis LGRB afterglow), which cannot be overcome
by their higher volumetric rates.  Also, as in the case of LLGRBs,
host galaxy contamination is a significant problem, especially for
SKA-Low.

Although it has been argued that perhaps half of all
core-collapse SNe are dust obscured \citep{Horiuchi+11}, the actual
fraction will need to be much higher than this ($\sim 90\%$, i.e., a
true core-collapse SN rate that is 10 times higher than inferred from
optical surveys) to allow a meaningful rate measurement from
untargeted radio surveys due to the small number of detectable events.
Because the low-luminosity events do not evolve rapidly (typical
timescales $\gtrsim 100$ days at 1.4 GHz) and are only detectable in
the local universe, targeted searches of local galaxies with a few
epochs per year are likely to be much more efficient for investigating
the prevalence of obscured event populations ($\S\ref{sec:RSNe}$).

\subsection{Summary}

\autoref{fig:yields} graphically summarizes the expected number of
event detections for each source class and survey.  We briefly
summarize the results of our simulations as follows:
\begin{itemize}

\item{No transients are expected to be detected with LOFAR (even the
    hypothetical LOFAR-Expanded survey) due to a mismatch between the
    survey sensitivity and the low peak luminosities of transients at
    sub-GHz frequencies (self-absorption).  The long durations at
    these frequencies also lead to many events appearing as steady
    sources.}

\item{At GHz frequencies, the wide surveys VAST-Wide-Stack, VAST-Deep,
    and VLASS-Wide are the only pre-SKA surveys producing more than a
    handful of detections, including $\sim 15-30$ classical LGRBs
    (mostly off-axis), $\sim 10-40$ TDEs (mostly off-axis), $\sim 2$
    on-axis SGRBs, and $\sim 5-20$ NSM-magnetar events.  The deep
    pencil-beam surveys may detect a few transients.  However, even
    though they are sensitive to events at higher redshifts, where the
    volumetric rate may be higher (e.g., in the case of LGRBs) and
    positive K-corrections become important, this is insufficient to
    overcome their lower \'etendue compared to the wide surveys.}

\item{The CMB-Stack surveys at 150 GHz detect only on-axis events:
    $\sim 40$ LGRBs and $\sim 2$ jetted TDEs.}

\item{The SKA surveys (in particular our hypothetical best-case
    SKA-Expanded survey) detect hundreds or thousands of LGRBs (on-
    and off-axis), off-axis jetted TDEs, and NSM-magnetar events.
    They also detect smaller numbers ($\sim 10-100$) of LLGRBs, SGRBs,
    and NS-NS mergers with prompt black hole formation.}

\item{Essentially none of the surveys (with the exception of a few
    detections with SKA-Expanded) detect Type Ib/c SNe due to their
    low luminosity.  While core-collapse SNe have been discovered in
    radio observations, these events are not efficiently discovered in
    untargeted surveys.}

\end{itemize}

\section{Conclusions}
\label{sec:discussion}

We presented general arguments about the constrained
timescale-luminosity relation of extragalactic synchrotron transients,
and utilized the light curves for a wide range of known and
hypothetical transients and the properties of multiple planned and
hypothetical radio surveys spanning 150 MHz to 150 GHz to explore the
detection rates and basic properties of radio transients.  Our finding
can be summarized as follows:

\begin{itemize}

\item The basic physics of synchrotron-emitting extragalactic
  transients place significant constraints on their possible radio
  light curves.  Of particular relevance to radio transient surveys
  are the facts that luminous events cannot evolve on arbitrarily
  short timescales, and that the emission evolves from high to low
  frequencies.  Events with radio luminosities sufficient to be
  detected in appreciable numbers with future radio surveys (e.g.,
  LGRBs, TDEs) have characteristic timescales of $\gtrsim 100$ days at
  GHz frequencies and about an order of magnitude longer at $\sim 0.1$
  GHz.

\item Our Monte Carlo simulations predict detection rates for planned
  and hypothetical radio transient surveys, for the first time using
  an end-to-end model including realistic volumetric rates and light
  curves, and simulated survey strategies.  We account for all
  relevant cosmological effects, including time dilation,
  K-corrections, rate evolution, and non-Euclidean luminosity
  distance, which are important in the context of the most sensitive
  surveys such as SKA.

\item Low-luminosity transients require volumetric rates
  comparable to the local core-collapse SN rate to be discoverable in
  untargeted radio transient surveys.  They will therefore account for
  at most a few events in all pre-SKA surveys, and even with SKA will
  number $\sim 10-100$ in the entire survey.  Such transients are
  therefore best studied through targeted radio follow-up of
  discoveries from other wavelengths, if they produce emission outside
  of the radio band.  However, even in the case that they produce only
  radio emission, or are dust-obscured, a more profitable approach may
  be to target nearby galaxies.

\item The small number of expected detections in the pre-SKA surveys
  (and the hundreds to thousands of detections with SKA) are dominated
  by relativistic sources (LGRBs, TDEs, NSM-magnetars) both on- and
  off-axis.  These events are detected at cosmological distances where
  redshift determinations from host galaxies may prove challenging,
  and the ability to locate them to specific regions of the hosts
  (e.g., the nucleus in the case of TDEs) may be limited.

\item Our hypothetical millimeter surveys will detect mainly on-axis LGRBs and TDEs.  The detection rates we calculate may be enhanced if additional events are detected by their reverse shock emission, which is significantly brighter than the forward shock emission but shorter lived.  LGRBs discovered by the stacked CMB surveys via their forward shock emission could also be searched at early survey epochs for reverse shock emission.

\item Our predicted detection rates are generally lower than previous
  claims (\autoref{fig:yields}).  We predict that the most successful
  pre-SKA survey will be VAST-Deep, with $\sim 70$ extragalactic
  transient detections across all considered classes.  In some cases,
  the lower numbers that we predict are due to more conservative
  assumptions about the underlying events (e.g.,
  \S\ref{sec:offaxistderesults}).  Another factor is our $10\sigma$
  detection criterion, which is generally higher than previous
  studies, but also more realistic given the scale of the surveys and
  lessons about false-positive candidates from past searches
  \citep{Frail+12}.  Our simulations also highlight the impact of slow
  evolutionary timescales at low frequencies which will restrict the
  ability to separate transient sources from variable and steady
  sources.  We also note that some previous studies have overestimated
  the effective areas of surveys that observe at cadences
  significantly higher than the relevant evolution times.

\item We find that host galaxy contamination is not expected to be a problem
  for most detected transients, because the detections are dominated by
  luminous events that outshine their host galaxies for moderate star
  formation rates of $\sim$1--10~M$_\odot$~yr$^{-1}$. The small numbers of
  detections of the low-luminosity transient classes will be reduced at
  $\lesssim$ GHz frequencies, however, if the host galaxy emission cannot be
  subtracted cleanly. This is especially relevant for SKA-Low, because at low
  frequency only a few radio transients (on-axis SGRBs, NSM-magnetars, TDEs)
  manage to significantly outshine their host galaxies, and the resolution of
  low frequency transient surveys is relatively poor ($11''$ for SKA-Low).

\item For near-term surveys, we find that areal densities of realistic
  extragalactic radio transients generally follow a Euclidean scaling
  relation, $N_{\rm all-sky}({>}F_\nu) \propto F_\nu^{-3/2}$, where $N_{\rm
    all-sky}({>}F_\nu)$ is number of transients on the sky brighter than some
  flux density cutoff, $F_\nu$ (\autoref{fig:lnls}). As a rule of thumb,
  non-Euclidean effects are not significant for transients brighter than
  $\sim$1~mJy in almost all circumstances, and for some sources they can be
  ignored down to flux densities of $\sim 10$ $\mu$Jy. However, for the more
  sensitive SKA surveys, which are the only surveys expected to detect a large
  number of transients, non-Euclidan effects and cosmological source evolution
  cannot be neglected.

\item The small number of predicted events in pre-SKA surveys will
  inhibit attempts to measure event rates and beaming fractions, since
  the uncertainties induced by small-number statistics will be
  significant.  It is also unclear if the various transient sources
  can be distinguished from each other in surveys that span only a
  small number of epochs.

\item The SKA will detect hundreds to thousands of events. These will be
  dominated by LGRBs but will also include SGRBs (primarily on-axis), off-axis
  TDEs, and NSM-magnetar events.

\item NSM-magnetar events have the potential to be a major source of
  radio transients and may dominate the yield of SKA-Low.  However,
  the underlying rates and light curves are currently highly
  uncertain; upcoming surveys will constrain them.

\end{itemize}

We have not considered the question of what astrophysical information
can be extracted once a radio transient is discovered.  As we have
emphasized, the general lag of radio emission behind that at other
bands makes the detailed characterization of radio-discovered
transients challenging. Paper II will investigate this topic in
detail.

\acknowledgements

This research has made use of the SIMBAD database, operated at CDS,
Strasbourg, France, and NASA's Astrophysics Data System. BDM acknowledges
helpful conversations with Dimitrios Giannios, Bradley Johnson, Glenn Jones, David Kaplan,
Peter Mimica, and Ashley Zauderer. We thank Aleksey Generozov for technical
assistance and Rodolfo Barniol-Duran for providing LLGRB light curve models.
BDM gratefully acknowledges support from NASA {\it Fermi} grant NNX14AQ68G,
NSF grant AST-1410950, and the Alfred P. Sloan Foundation.

\begin{deluxetable*}{lcccccccc|ccccccccc}
\tabletypesize{\scriptsize}
\tablecolumns{15}
\tabcolsep0.025in\footnotesize
\tablewidth{0pc}
\tablecaption{LGRBs
\label{tab:LGRB}}
\tablehead {
\colhead{Survey} &
\colhead{$N^{a}$} &
\colhead{$\bar{n}^{b}$} &
\colhead{$\bar{z}^{c}$} &
\colhead{$f_{\rm V}^{d}$} &
\colhead{$f_{\rm rise}^{e}$} &
\colhead{$f_{\rm fall}^{f}$} &
\colhead{$f_{\rm peak}^{g}$} &
\colhead{gal?$^{h}$} &
\colhead{$N$} &
\colhead{$\bar{n}$} &
\colhead{$\bar{z}$} &
\colhead{$f_{\rm V}$} &
\colhead{$f_{\rm rise}$} &
\colhead{$f_{\rm fall}$} &
\colhead{$f_{\rm peak}$} &
\colhead{gal?}
}
\startdata
 & \multicolumn{7}{c}{$\theta_{\rm obs} = 0.2$} & \multicolumn{7}{c}{$\theta_{\rm obs} = 0.4$} \\
\hline
LOFAR &  0.0 &  26(37) & 0.004 & 0.48 & 0.75 & 0.25 & 0.58 & M &0.0 &  25(37) & 0.004 & 0.48 & 0.82 & 0.18 & 0.61& M \\
LOFAR-E &  0.0 &  26(37) & 0.012 & 0.46 & 0.77 & 0.23 & 0.61 & M & 0.0 &  25(37) & 0.012 & 0.49 & 0.82 &0.18 & 0.60& M \\
SKA-Low &  5.1 &   10(13) & 0.249 & 0.28 & 0.68 & 0.54 & 0.64 & M &  12 &   9(13) & 0.243 & 0.38 & 0.75 &0.28 & 0.66& M \\
VAST-W &  0.09 & 217(1096) & 0.074 & 0.00 & 0.91 & 0.90 & 0.91 & N & 0.18 & 277(1096) & 0.062 & 0.00 & 0.87 &0.87 & 0.88& N \\
VAST-W-S &  1.4 &   7(37) & 0.184 & 0.00 & 0.91 & 0.90 & 0.88 & N & 2.6 &   10(37) & 0.152 & 0.00 & 0.87 &0.87 & 0.85& N \\
VAST-D &  1.4 &   1.1(4) & 0.219 & 0.01 & 0.73 & 0.76 & 0.00 & N & 3.6 &   1.2(4) & 0.184 & 0.01 & 0.75 &0.74 & 0.00& N \\
VAST-D-SF &  0.0 & 170(366) & 0.252 & 0.00 & 0.75 & 0.76 & 0.78 &N& 0.010 & 208(366) & 0.203 & 0.00 & 0.68 &0.68) & 0.71& N \\
VAST-D-SF-S &  0.16 &   6(13) & 0.847 & 0.00 & 0.75 &0.73 & 0.73 &N &  0.16 &   8(13) & 0.545 & 0.00 & 0.67 &0.66 & 0.65& N \\
SKA & 140 &   3(13) & 0.914 & 0.00 & 0.89 & 0.90 & 0.31 &N & 150 &   3(13) & 0.589 & 0.00 & 0.85 &0.84 & 0.64& N \\
SKA-E & 2060 &   8(37) & 1.822 & 0.00 & 0.89 & 0.87 & 0.87 &N & 1300 &  10(37) & 0.826 & 0.00 & 0.85 &0.84 & 0.84& N \\
VLASS-W &  1.9 &   1.0(4) & 0.310 & 0.00 & 0.76 & 0.75 & 0.00 & N &  4.2 &   1.0(4) & 0.223 & 0.00 & 0.75 &0.75 & 0.00& N \\
VLASS-D &  0.45 &   1.0(4) & 1.892 & 0.00 & 0.76 & 0.75 & 0.00 &N& 2.0 &   1.1(4) & 2.066 & 0.00 & 0.75 &0.74 & 0.00& N \\
CMB &  8.2 &  25(1096) & 0.343 & 0.00 & 0.99 & 0.99 & 0.99 &N & 1.1 &  61(1096) & 0.117 & 0.00 & 0.98 &0.97 & 0.98& N \\
CMB-S1 & 36 &   2.9(110) & 0.568 & 0.00 & 0.99 & 0.99 & 0.05 &N&  5.6 &   7(110) & 0.199 & 0.00 & 0.97 &0.97 & 0.58& N \\
CMB-S2 & 32 &   1.0(11) & 0.847 & 0.00 & 0.90&0.91 & 0.00 &N & 8.4 &   1(11) & 0.282 & 0.00 & 0.90 &0.91 & 0.00& N \\
\hline
 & \multicolumn{7}{c}{$\theta_{\rm obs} = 0.8$} & \multicolumn{7}{c}{$\theta_{\rm obs} = 1.57$} \\
\hline
LOFAR & 0.0 &  29(37) & 0.004 & 0.44 & 0.60&0.40 & 0.34& M  & 0.0 &  32(37) & 0.004 & 0.18 & 0.50 &0.50 & 0.37& M \\
LOFAR-E &  0.008 &  29(37) & 0.011 & 0.44 & 0.61&0.39 & 0.36 & M &0.04 &  32(37) & 0.014 & 0.19 & 0.46 &0.54 & 0.34& M \\
SKA-Low &  45 &   10(13) & 0.203 & 0.49 & 0.72&0.28 & 0.43& M  & 320 &  12(13) & 0.252 & 0.23 & 0.47 &0.53 & 0.28& M \\
VAST-W &  0.40 & 439(1096) & 0.050 & 0.00 & 0.79&0.82 & 0.86 & N & 0.42 & 743(1096) & 0.032 & 0.00 & 0.62 &0.64 & 0.75& N \\
 VAST-W-S & 6.2 &  15(37) & 0.121 & 0.00 & 0.78&0.81 & 0.78& N  & 5.4 &  26(37) & 0.073 & 0.00 & 0.57 &0.63 & 0.60& N \\
 VAST-D & 11.8 &   1.5(4) & 0.149 & 0.00 & 0.73&0.74 & 0.01 & N & 13.0 &   2.5(4) & 0.097 & 0.00 & 0.62 &0.57 & 0.25& N \\
 VAST-D-SF &  0.50 & 435(1096) & 0.166 & 0.00 & 0.79&0.81 & 0.85 & N &  0.04 & 769(1096) & 0.099 & 0.00 & 0.60 &0.65 & 0.75& N \\
VAST-D-SF-S &  0.68 &  15(37) & 0.404 & 0.00 & 0.79&0.78 & 0.78 & N &0.42 &  26(37) & 0.205 & 0.00 & 0.58 &0.58) & 0.57& N \\
SKA & 270  &   5(13) & 0.421 & 0.00 & 0.77&0.78 & 0.71 & N & 160 &   9(13) & 0.209 & 0.00 & 0.58 &0.59 & 0.52& N \\
SKA-E &1900 &  16(37) & 0.567 & 0.00 & 0.78&0.78 & 0.79& N & 860 &  27(37) & 0.254 & 0.00 & 0.55 &0.60 & 0.54& N \\
 VLASS-W & 7.6 &   1.1(4) & 0.141 & 0.00 & 0.75&0.76 & 0.00& N  & 2.2 &   2.4(4) & 0.054 & 0.00 & 0.63 &0.59 & 0.19& N \\
 VLASS-D & 2.3 &   1.3(4) & 0.887 & 0.00 & 0.75&0.75 & 0.00 & N & 0.30 &   2.8(4) & 0.256 & 0.00 & 0.57 &0.55 & 0.23& N \\
CMB &   0.018 & 236(1096) & 0.018 & 0.00 & 0.88&0.87 & 0.89 & N & 0.0 & 727(1096) & 0.004 & 0.00 & 0.63 &0.63 & 0.74& N \\
CMB-S1 &  0.10 &  24(110) & 0.033 & 0.00 & 0.89&0.87 & 0.88 & N &  0.06 &  72(110) & 0.007 & 0.00 & 0.62 &0.62 & 0.63& N \\
CMB-S2 &  0.44 &   3(11) & 0.054 & 0.00 & 0.87&0.86 & 0.16 & N &  0.03 &   8(11) & 0.013 & 0.00 & 0.61 &0.60 & 0.58& N
\enddata
\tablecomments{$^{a}$ Total number of detected events. $^{b}$ Average
  number of detected epochs (total epochs). $^{c}$ Average redshift of
  detected events. $^{d}$ Fraction of total detected events discarded
  for lack of variability. $^{e}$ Fraction of events with measured
  rise times. $^{f}$ Fraction of events with measured
  fall times. $^{g}$ Fraction of events with a measured peak.
  $^h$ Is galaxy contamination an issue?  (Yes, No, Maybe).}
\end{deluxetable*}

\begin{deluxetable*}{lcccccccc}
\tabletypesize{\scriptsize}
\tablecolumns{8}
\tabcolsep0.025in\footnotesize
\tablewidth{0pc}
\tablecaption{LLGRBs
\label{tab:LLGRB}}
\tablehead {
\colhead{Survey} &
\colhead{$N$} &
\colhead{$\bar{n}$} &
\colhead{$\bar{z}$} &
\colhead{$f_{\rm V}$} &
\colhead{$f_{\rm rise}$} &
\colhead{$f_{\rm fall}$} &
\colhead{$f_{\rm peak}$} &
\colhead{gal?}
}
\startdata
LOFAR & 0.0 &  17(37) & 0.001 & 0.00 & 0.82 & 0.86 & 0.81 & Y \\
LOFAR-E & 0.0 &  17(37) & 0.002 & 0.00 & 0.81 & 0.85 & 0.80 & Y \\
SKA-Low & 18 &   6(13) & 0.036 & 0.00 & 0.81 & 0.83 & 0.68 & Y \\
VAST-W & 0.40 & 140(1096) & 0.005 & 0.00 & 0.96 & 0.96 & 0.96 & Y \\
VAST-W-S & 0.36 &   6(37) & 0.011 & 0.00 & 0.95&0.95 & 0.30 & Y \\
VAST-D & 0.18 &   1.2(4) & 0.012 & 0.00 & 0.73&0.76 & 0.00 & Y \\
VAST-D-SF &  0.0 & 141(1096) & 0.016 & 0.00 & 0.96&0.97 & 0.96 & Y \\
VAST-D-SF-S &   0.03 &   6(37) & 0.031 & 0.00 & 0.95&0.94 & 0.29 & Y \\
SKA & 6.6 &   2(13) & 0.029 & 0.00 & 0.90&0.93 & 0.02 & M \\
SKA-E& 64 &   6(37) & 0.040 & 0.00 & 0.94&0.95 & 0.30 & M \\
VLASS-W & 0.060 &   1.1(4) & 0.010 & 0.00 & 0.77&0.72 & 0.00& M \\
VLASS-D & 0.012 &   1.1(4) & 0.053 & 0.00 & 0.74&0.77 & 0.00& Y \\
CMB&  0.04 &   4(1096) & 0.005 & 0.00 & 1.00& 1.00 & 0.00 & Y \\
CMB-S1& 0.034 &   1(110) & 0.008 & 0.00 & 1.00&1.00 & 0.00 & Y \\
CMB-S2& 0.016 &   1(11) & 0.012 & 0.00 & 0.89& 0.89 & 0.00& Y
\enddata
\tablecomments{Columns are as in \autoref{tab:LGRB}.}
\end{deluxetable*}

\begin{deluxetable*}{lcccccccc|ccccccccc}
\tabletypesize{\scriptsize}
\tablecolumns{15}
\tabcolsep0.025in\footnotesize
\tablewidth{0pc}
\tablecaption{SGRBs
\label{tab:SGRB}}
\tablehead {
\colhead{Survey} &
\colhead{$N$} &
\colhead{$\bar{n}$} &
\colhead{$\bar{z}$} &
\colhead{$f_{\rm V}$} &
\colhead{$f_{\rm rise}$} &
\colhead{$f_{\rm fall}$} &
\colhead{$f_{\rm peak}$} &
\colhead{gal?} &
\colhead{$N$} &
\colhead{$\bar{n}$} &
\colhead{$\bar{z}$} &
\colhead{$f_{\rm V}$} &
\colhead{$f_{\rm rise}$} &
\colhead{$f_{\rm fall}$} &
\colhead{$f_{\rm peak}$} &
\colhead{gal?}
}
\startdata
 & \multicolumn{7}{c}{$\theta_{\rm obs} = 0.2$} & \multicolumn{7}{c}{$\theta_{\rm obs} = 0.4$} \\
\hline
LOFAR &0.0 &  12(37) & 0.003 & 0.00 & 0.83&0.83 & 0.82 & N& 0.0&  16(37) & 0.002 & 0.00 & 0.77&0.76 & 0.76& M\\
 LOFAR-E & 0.0 &  12(37) & 0.010 & 0.00 & 0.83&0.85 & 0.83 & N&0.006 &  16(37) & 0.008 & 0.00 & 0.75&0.76 & 0.73& M\\
 SKA-Low & 22 &   4(13) & 0.173 & 0.00 & 0.81&0.83 & 0.53 & N&28 &   6(13) & 0.125 & 0.00 & 0.73&0.75 & 0.69& M\\
VAST-W &  0.14 & 153(1096) & 0.034 & 0.00 & 0.91&0.92 & 0.92 & N&0.06 & 270(1096) & 0.018 & 0.00 & 0.85&0.86 & 0.86& N\\
 VAST-W-S & 1.8 &   5(37) & 0.082 & 0.00 & 0.91&0.91 & 0.83 & N&0.90 &   9(37) & 0.043 & 0.00 & 0.86&0.83 & 0.84& N\\
 VAST-D & 1.5 &   1.0(4) & 0.098 & 0.00 & 0.74&0.74) & 0.00& N & 1.1 &   1.1(4) & 0.051 & 0.00 & 0.75&0.75 & 0.00& N\\
VAST-D-SF &   0.016 & 160(1096) & 0.110 & 0.00 & 0.91&0.92 & 0.92 & N &0.006 & 276(1096) & 0.057 & 0.00 & 0.85&0.85 & 0.86& N\\
 VAST-D-SF-S &  0.20 &   6(37) & 0.255 & 0.00 & 0.91&0.90 & 0.89 & N &0.078 &   10(37) & 0.129 & 0.00 & 0.85& 0.85 & 0.83& N\\
 SKA &  66 &   2.1(13) & 0.258 & 0.00 & 0.89&0.90 & 0.04 & N&30 &   3(13) & 0.135 & 0.00 & 0.84&0.84 & 0.45& N\\
 SKA-E & 420 &   6.1(37) & 0.335 & 0.00 & 0.90&0.91 & 0.89 & N&170 &   10(37) & 0.169 & 0.00 & 0.85&0.84 & 0.83& N\\
VLASS-W &  1.2 &   1.0(4) & 0.097 & 0.00 & 0.77&0.75 & 0.00& N &0.64 &   1.0(4) & 0.044 & 0.00 & 0.77&0.76 & 0.00& N\\
VLASS-D &  0.6 &   1.0(4) & 0.707 & 0.00 & 0.77&0.74 & 0.00 & N& 0.18 &   1.0( 4) & 0.264 & 0.00 & 0.76&0.74 & 0.00& N\\
CMB &   0.0 & 150(1096) & 0.004 & 0.00 & 0.92&0.92 & 0.92 & N&  0.0 & 270(1096) & 0.002 & 0.00 & 0.86&0.87 & 0.87& N\\
CMB-S1 &   0.0 &  15(110) & 0.007 & 0.00 & 0.92&0.92 & 0.92 & N& 0.0 &  27(110) & 0.004 & 0.00 & 0.86&0.85 & 0.86& N\\
CMB-S2 &  0.006 &   2(11) & 0.012 & 0.00 & 0.89&0.91 & 0.00 & N&0.0 &   3(11) & 0.007 & 0.00 & 0.84&0.84 & 0.29& N\\
\hline
 & \multicolumn{7}{c}{$\theta_{\rm obs} = 0.8$} & \multicolumn{7}{c}{$\theta_{\rm obs} = 1.57$} \\
\hline
LOFAR &  0.0 &  25(37) & 0.001 & 0.01 & 0.58&0.61 & 0.62 & M&0.0 &  33(37) & 0.001 & 0.39 & 0.57&0.43 & 0.19& M\\
LOFAR-E &  0.006 &  25(37) & 0.004 & 0.01 & 0.60&0.59 & 0.60 & M&0.0 &  33(37) & 0.002 & 0.38 & 0.50&0.50 & 0.15& M\\
SKA-Low & 28 &   9(13) & 0.070 & 0.01 & 0.57&0.58 & 0.59 & M& 7.0 &  12(13) & 0.029 & 0.41 & 0.55&0.45 & 0.14& M\\
VAST-W &  0.008 & 587(1096) & 0.005 & 0.00 & 0.69&0.70 & 0.76 & M&0.0 & 960(1096) & 0.001 & 0.38 & 0.55&0.45 & 0.53& M\\
 VAST-W-S & 0.11 &  20(37) & 0.012 & 0.00 & 0.69&0.70 & 0.70 & M& 0.006 &  32(37) & 0.003 & 0.41 & 0.55&0.45 & 0.22& M\\
VAST-D &  0.24 &   2.0(4) & 0.016 & 0.00 & 0.67&0.68 & 0.09 & M& 0.016 &   3.4(4) & 0.004 & 0.41 & 0.53&0.47 & 0.11& M\\
VAST-D-SF &   0.0 & 601(1096) & 0.017 & 0.00 & 0.70&0.72 & 0.76& M &0.0 & 950(1096) & 0.004 & 0.41 & 0.53&0.47 & 0.49& M\\
VAST-D-SF-S &  0.010 &  20(37) & 0.036 & 0.00 & 0.70&0.66 & 0.66 & M&0.0 &  32(37) & 0.008 & 0.43 & 0.56&0.44 & 0.23& M\\
 SKA &  3.5 &   7(13) & 0.038 & 0.00 & 0.68&0.69 & 0.63 & M&0.20 &  11(13) & 0.009 & 0.42 & 0.61&0.39 & 0.20& M\\
 SKA-E &  20 &  20(37) & 0.048 & 0.00 & 0.69&0.70 & 0.69 & M&1.0 &  33(37) & 0.011 & 0.44 & 0.55&0.45 & 0.26& M\\
VLASS-W &   0.072 &   1.9(4) & 0.011 & 0.00 & 0.68&0.71 & 0.04 & M& 0.002 &   3.4(4) & 0.002 & 0.41 & 0.59&0.41 & 0.15& M\\
VLASS-D &   0.014 &   1.9(4) & 0.062 & 0.00 & 0.71&0.72 & 0.06 & M& 0.0 &   3.4(4) & 0.012 & 0.41 & 0.57&0.43 & 0.15& M\\
CMB &   0.0 & 591(1096) & 0.001 & 0.00 & 0.72&0.71 & 0.77& N &0.0 & 825(1096) & 0.000 & 0.00 & 0.64&0.80 & 0.78& N\\
CMB-S1 &   0.0 &  60(110) & 0.001 & 0.00 & 0.68&0.73 & 0.72 & N&  0.0 &  82(110) & 0.001 & 0.00 & 0.65&0.78 & 0.75& N\\
 CMB-S2 &   0.0 &   6(11) & 0.002 & 0.00 & 0.67&0.67 & 0.61 & N&0.0 &   8(11) & 0.001 & 0.06 & 0.65&0.64 & 0.65& N\\
\enddata
\tablecomments{Columns are as in \autoref{tab:LGRB}.}
\end{deluxetable*}

\begin{deluxetable*}{lcccccccc}
\tabletypesize{\scriptsize}
\tablecolumns{8}
\tabcolsep0.025in\footnotesize
\tablewidth{0pc}
\tablecaption{NS-NS Mergers with Prompt Black Hole Formation
\label{tab:NSM}}
\tablehead {
\colhead{Survey} &
\colhead{$N$} &
\colhead{$\bar{n}$} &
\colhead{$\bar{z}$} &
\colhead{$f_{\rm V}$} &
\colhead{$f_{\rm rise}$} &
\colhead{$f_{\rm fall}$} &
\colhead{$f_{\rm peak}$} &
\colhead{gal?}
}
\startdata
LOFAR & 0.0 &  36(37) & 0.001 & 0.82 & 1.00 & 0.00 & 0.23 & M\\
LOFAR-E &  0.016 &  37(37) & 0.003 & 0.81 & 1.00 & 0.00 & 0.21 & M\\
SKA-Low & 48 &  13(13) & 0.046 & 0.84 & 1.00 & 0.00 & 0.04 & M\\
VAST-W & 0.0 & 1095(1096) & 0.002 & 0.87 & 1.00 & 0.00 & 0.33 & M\\
VAST-W-S &  0.060 &  37(37) & 0.005 & 0.86 & 1.00 & 0.00 & 0.15 & M\\
VAST-D &  0.13 &   3.9(4) & 0.008 & 0.86 & 1.00 & 0.00 & 0.00 & M\\
VAST-D-SF &  0.0 & 1095(1096) & 0.007 & 0.80 & 1.00 & 0.00 & 0.30& M \\
 VAST-D-SF-S & 0.006 &  37(37) & 0.015 & 0.84 & 1.00 & 0.00 & 0.13 & M\\
 SKA & 1.9 &  13(13) & 0.018 & 0.84 & 1.00 & 0.00 & 0.08 & M\\
SKA-E &  12 &  37(37) & 0.020 & 0.82 & 1.00 & 0.00 & 0.09 & M\\
 VLASS-W &  0.026 &   4.0(4) & 0.004 & 0.83 & 1.00 & 0.00 & 0.00 & N\\
VLASS-D &   0.0 &   3.9(4) & 0.021 & 0.86 & 1.00 & 0.00 & 0.00 & N\\
CMB &   0.0 & 1096(1096) & 0.000 & 0.80 & 1.00 & 0.00 & 0.48 & N\\
CMB-S1 &   0.0 & 110(110) & 0.001 & 0.83 & 1.00 & 0.00 & 0.20 & N\\
CMB-S2 &   0.0 &  11(11) & 0.001 & 0.86 & 1.00 & 0.00 & 0.02 & N
\enddata
\tablecomments{Columns are as in \autoref{tab:LGRB}.}
\end{deluxetable*}

\begin{deluxetable*}{lcccccccc}
\tabletypesize{\scriptsize}
\tablecolumns{8}
\tabcolsep0.025in\footnotesize
\tablewidth{0pc}
\tablecaption{NS-NS Mergers with a Stable Magnetar Remnant
\label{tab:NSM-magnetar}}
\tablehead {
\colhead{Survey} &
\colhead{$N$} &
\colhead{$\bar{n}$} &
\colhead{$\bar{z}$} &
\colhead{$f_{\rm V}$} &
\colhead{$f_{\rm rise}$} &
\colhead{$f_{\rm fall}$} &
\colhead{$f_{\rm peak}$} &
\colhead{gal?}
}
\startdata
LOFAR & 0.0 &  33(37) & 0.038 & 0.76 & 1.00 & 0.00 & 0.00 & N\\
LOFAR-E &  0.90 &  33(37) & 0.131 & 0.76 & 1.00 & 0.00 & 0.14& N\\
SKA-Low & 1160 &  12(13) & 1.425 & 0.86 & 1.00 & 0.00 & 0.15& N\\
VAST-W & 1.2 & 870(1096) & 0.140 & 0.58 & 0.95 & 0.05 & 0.73& N\\
VAST-W-S &  12 &  31(37) & 0.296 & 0.65 & 1.00 & 0.00 & 0.61& N\\
VAST-D &  22 &   3.2(4) & 0.371 & 0.70 & 1.00 & 0.00 & 0.31& N\\
VAST-D-SF &  0.080 & 945(1096) & 0.388 & 0.65 & 0.99&0.01 & 0.82& N\\
 VAST-D-SF-S & 0.72 &  33(37) & 0.755 & 0.72 & 1.00 & 0.00 & 0.53& N\\
 SKA & 220 &  12(13) & 0.805 & 0.76 & 1.00 & 0.00 & 0.50& N\\
SKA-E &  1020 &  34(37) & 0.887 & 0.75 & 1.00 & 0.00 & 0.40& N\\
 VLASS-W &  5.0 &   3.1(4) & 0.227 & 0.63 & 0.98&0.02 & 0.39& N\\
VLASS-D &   0.38 &   3.5(4) & 0.948 & 0.76 & 1.00 & 0.00 & 0.16& N\\
CMB &   0.006 & 917(1096) & 0.024 & 0.47 & 0.77&0.23 & 0.72& N\\
CMB-S1 &  0.032 &  92(110) & 0.042 & 0.47 & 0.79&0.21 & 0.67& N\\
CMB-S2 & 0.13 &   9(11) & 0.070 & 0.60 & 1.00 & 0.00 & 0.62& N
\enddata
\tablecomments{Columns are as in \autoref{tab:LGRB}.}
\end{deluxetable*}

\begin{deluxetable*}{lcccccccc|ccccccccc}
\tabletypesize{\scriptsize}
\tablecolumns{15}
\tabcolsep0.025in\footnotesize
\tablewidth{0pc}
\tablecaption{TDEs
\label{tab:TDE}}
\tablehead {
\colhead{Survey} &
\colhead{$N$} &
\colhead{$\bar{n}$} &
\colhead{$\bar{z}$} &
\colhead{$f_{\rm V}$} &
\colhead{$f_{\rm rise}$} &
\colhead{$f_{\rm fall}$} &
\colhead{$f_{\rm peak}$} &
\colhead{gal?} &
\colhead{$N$} &
\colhead{$\bar{n}$} &
\colhead{$\bar{z}$} &
\colhead{$f_{\rm V}$} &
\colhead{$f_{\rm rise}$} &
\colhead{$f_{\rm fall}$} &
\colhead{$f_{\rm peak}$} &
\colhead{gal?}
}
\startdata
 & \multicolumn{7}{c}{\swsixteen\ (On-Axis)} & \multicolumn{7}{c}{Off-Axis} \\
\hline
LOFAR & 0.0 &  27(37) & 0.013 & 0.00 & 0.62&0.64 & 0.62 & N& 0.0 &  29(37) & 0.021 & 0.90 & 1.00 & 0.00 & 0.28 & N\\
LOFAR-E &  0.0 &  27(37) & 0.042 & 0.00 & 0.59&0.63 & 0.59 & N& 0.20 &  30(37) & 0.072 & 0.91 & 1.00 & 0.00 & 0.28& N\\
SKA-Low &  1.9 &  11(13) & 0.744 & 0.14 & 0.48&0.52 & 0.35 & N& 280 &  11(13) & 1.271 & 0.84 & 1.00 & 0.00 & 0.20& N\\
VAST-W &  0.062 & 871(1096) & 0.175 & 0.02 & 0.50&0.51 & 0.52 & N& 2.0 & 902(1096) & 0.129 & 0.28 & 0.64&0.42 & 0.63& N\\
VAST-W-S &  0.52 &  29(37) & 0.424 & 0.27 & 0.58&0.42 & 0.45 & N& 13 &  31(37) & 0.275 & 0.39 & 0.66&0.35 & 0.63& N\\
VAST-D & 0.82 &   2.9(4) & 0.561 & 0.46 & 0.71&0.29 & 0.31 & N&  20 &   3.2(4) & 0.332 & 0.54 & 0.74&0.26 & 0.38& N\\
VAST-D-SF &  0.0 & 289(366) & 0.595 & 0.82 & 1.00 & 0.00 & 0.07 & N&  0.018 & 359(366) & 0.320 & 0.82 & 1.00 & 0.00 & 0.23& N\\
VAST-D-SF-S & 0.0 &   10(13) & 1.542 & 0.90 & 1.00 & 0.00 & 0.00 & N& 0.070 &  13(13) & 0.507 & 0.88 & 1.00 & 0.00 & 0.08& N\\
SKA & 2.6 &   8(13) & 1.528 & 0.76 & 0.95&0.05 & 0.41 & N& 72 &  11(13) & 0.624 & 0.72 & 0.90&0.10 & 0.50& N\\
SKA-E & 8.4 &  22(37) & 1.641 & 0.80 & 0.94&0.06 & 0.40 & N& 240 &  31(37) & 0.713 & 0.71 & 0.90&0.10 & 0.53& N\\
VLASS-W & 0.80 &   2.3(4) & 0.719 & 0.54 & 0.94&0.06 & 0.25 & N& 6.0 &   3.2(4) & 0.202 & 0.38 & 0.68&0.33 & 0.34& N\\
VLASS-D & 0.0 &   2.1(4) & 1.697 & 0.74 & 0.98&0.02 & 0.27 & N& 0.11 &   3.3(4) & 0.742 & 0.74 & 0.91&0.09 & 0.32& N\\
CMB & 0.22 & 580(1096) & 0.359 & 0.00 & 0.82&0.92 & 0.82 & N& 0.012 & 903(1096) & 0.021 & 0.18 & 0.65&0.52) & 0.66& N\\
CMB-S1 & 0.86 &  60(110) & 0.649 & 0.00 & 0.80&0.93 & 0.79 & N& 0.060 &  89(110) & 0.038 & 0.22 & 0.64&0.47 & 0.64& N\\
CMB-S2 & 2.0 &   7(11) & 1.036 & 0.00 & 0.69&0.87 & 0.02 & N& 0.24 &   9(11) & 0.065 & 0.28 & 0.59&0.42 & 0.57& N
\enddata
\tablecomments{Columns are as in \autoref{tab:LGRB}.}
\end{deluxetable*}

\begin{deluxetable*}{lcccccccc}
\tabletypesize{\scriptsize}
\tablecolumns{8}
\tabcolsep0.025in\footnotesize
\tablewidth{0pc}
\tablecaption{Type Ib/c SNe
\label{tab:SNe}}
\tablehead {
\colhead{Survey} &
\colhead{$N$} &
\colhead{$\bar{n}$} &
\colhead{$\bar{z}$} &
\colhead{$f_{\rm V}$} &
\colhead{$f_{\rm rise}$} &
\colhead{$f_{\rm fall}$} &
\colhead{$f_{\rm peak}$} &
\colhead{gal?}
}
\startdata
LOFAR &0.0 &  30(37) & 0.001 & 0.07 & 0.59  & 0.54 & 0.60& Y\\
 LOFAR-E &0.0 &  30(37) & 0.002 & 0.12 & 0.60 &0.54 & 0.62& Y\\
 SKA-Low & 1.6 &  11(13) & 0.007 & 0.13 & 0.60 &0.53 & 0.59& Y\\
 VAST-W & 0.0 & 386(1096) & 0.001 & 0.00 & 0.88 &0.91 & 0.88& Y\\
 VAST-W-S & 0.032 &  13(37) & 0.002 & 0.00 & 0.88 &0.90 & 0.86& Y\\
 VAST-D & 0.036 &   1.8(4) & 0.002 & 0.00 & 0.71 &0.74 & 0.01& Y\\
 VAST-D-SF &  0.0 & 390(1096) & 0.003 & 0.00 & 0.88 &0.91 & 0.88& Y\\
VAST-D-SF-S & 0.0 &  14(37) & 0.006 & 0.00 & 0.87 &0.92 & 0.87& Y\\
SKA & 0.85 &   5(  13) & 0.006 & 0.00 & 0.84 &0.88 & 0.27& M\\
SKA-E &1.6 &  13(  37) & 0.008 & 0.00 & 0.87 &0.90 & 0.85& M\\
 VLASS-W & 0.010 &   1.4(4) & 0.002 & 0.00 & 0.71 &0.78 & 0.00& M\\
 VLASS-D & 0.0 &   1.5(4) & 0.010 & 0.00 & 0.72 &0.78 & 0.00& Y\\
CMB&  0.008 &  31(1096) & 0.001 & 0.00 & 1.00 &1.00 & 0.22& Y\\
 CMB-S1& 0.008 &  12(110) & 0.002 & 0.00 & 0.97 &0.98 & 0.19& Y\\
 CMB-S2& 0.006 &   3(11) & 0.003 & 0.06 & 0.65 &0.93 & 0.00& Y
\enddata
\tablecomments{Columns are as in \autoref{tab:LGRB}.}
\end{deluxetable*}

\clearpage

\appendix
\section{Limits from Previous Surveys}
\label{sec:limits}

Previous surveys are placed in Figure \ref{fig:lnls} according $F_{\nu,\rm
  lim}$, to their source detection thresholds\footnote{Note that in these
  works it is not necessarily true that $F_{\nu,\rm lim} = 10\sigma$, as we
  have adopted.}, and their constraints of N$_{\rm all-sky}$. Below we give
sensitivity numbers for these surveys. In the vast majority of cases no
sources are detected, and we take the 95 per cent Poisson credible interval on
N$_{\rm all-sky}$ given this fact. That limit is 3 sources in the survey area
\citep{Frail+12}, resulting in a lower limit of
\begin{equation}
N_{\rm all-sky} < 123800 \left(\frac{\Omega_s}{\rm deg^2}\right)^{-1},
\end{equation}
where $\Omega_s$ is the effective survey area, which is a function of the
transient evolutionary time scale $t_{\rm dur}$ because repeated visits to a
field do not add more information if the time between them is much shorter
than $t_{\rm dur}$ \citep{Macquart14}. We make the approximation that when new
epochs are obtained at a cadence $T_{\rm sep} \lesssim t_{\rm dur}$, the
effective survey area is
\begin{equation}
\Omega_s = \Omega_{\rm fp} \frac{T_{\rm bl}}{t_{\rm dur}},
\label{e.omegalarge}
\end{equation}
where $\Omega_{\rm fp}$ is footprint area of the survey and $T_{\rm bl}$ is
the total time baseline of the survey; the variability loss factor $V \approx
t_{\rm dur} / T_{\rm bl}$. A survey with a duration small compared to $t_{\rm
  dur}$ has $\Omega_s \rightarrow 0$, since separate observations result in
zero independent measurements in this case.

Based on the analysis in \autoref{sec:classes}, here we assume a 100-day
timescale in all cases. This means that in many cases our adopted values of
$\Omega_s$ are significantly smaller than the limiting case of the ``two-epoch
effective survey area'' \citep{Bower+07}:
\begin{equation}
\Omega_{s,{\rm 2E}} = \Omega_{\rm fp} (N_{\rm ep} - 1),
\label{e.omegasmall}
\end{equation}
where $N_{\rm ep}$ is the number of survey epochs, which applies only when
$T_{\rm sep} \gtrsim t_{\rm dur}$. In the case of ASKAP-Wide, $\Omega_s$ as
estimated this way is more than two orders of magnitude too large. We note
that because both Equations~\ref{e.omegalarge} and \ref{e.omegasmall} have
$\Omega_s \propto \Omega_{\rm fp} \propto \tau^{-1}$, however, wide and
shallow surveys are preferred for Euclidean sources regardless of the relevant
time scales.

\subsection{VLSS (150 MHz)}

\citet{Jaeger+12a} summarize ongoing work to perform a systematic
search for transients in the VLA Low Frequency Sky Survey (VLSS;
\citealt{Lane+12}). \citet{Jaeger+12} report one detection, which
occurred in one epoch of a 6-epoch search for transients in a 6.5
deg$^{2}$ field of view. The detection at ~5.5$\sigma$ (2.1 mJy)
occurred in the middle of the campaign, which lasted only 3
months. The source appears to rise in flux by a factor of two on a
~6-hr time scale in the detection epoch. Upper limits for the other
epochs are not provided by \citet{Jaeger+12}.

As the time scale of this survey is about the same as our 100-day choice, we
set the detection density to less than one source per 6.5 deg$^2$,
corresponding to $N_{\rm all-sky} <$ 6400. Note that this is well above the
predictions for most of our transient classes. The rapid variability observed
in the discovery epoch does not necessarily indicate unrealistically fast
timescales, as this may be symptomatic of the low significance of the
detection. To summarize, we take $F_{\nu,\rm lim} = 2.1$~mJy and $N_{\rm
  all-sky} < 6400$.

\subsection{FIRST (1.4 GHz)}

No transients are reported in comparison of the zero-declination strip of the
FIRST survey by \citet{deVries+04}, which was visited in two epochs separated
by 7 years. The quoted area is 120.2 deg$^2$ (their section 2) and the source
flux density limit is 2~mJy (their section 2.1.1). We take $F_{\nu,\rm lim} =
2$~mJy and $N_{\rm all-sky} < 1030$.

\subsection{ATATS/NVSS (1.4 GHz)}

\citet{Croft+10} compare the Allen Telescope Array Twenty-centimeter
Survey (ATATS) with NVSS, reporting no transients with a source flux
density limit of 40 mJy. As the time baseline between the two surveys
is $\gg 100$ days, the effective area is the ATATS footprint of 690
deg$^2$. We thus take $F_{\nu,\rm lim} = 40$~mJy and $N_{\rm all-sky} <
179$.

\subsection{FIRST/NVSS (1.4 GHz)}

\citet{Levinson+02} searched for transients in the overlap of the FIRST and
NVSS surveys. The effective search area is 2400 deg$^2$, and the
characteristic limit is 6 mJy. Since FIRST and NVSS were conducted
quasi-simultaneously, that the time baseline between the two epochs varied
between $\sim$0.5 and 4.5 yr. \citet{Gal-Yam+06} followed up the transient
candidates from \citet{Levinson+02}, concluding that two were real: one radio
supernovae and a second object without an apparent host galaxy. Based on our
calculations, the RSN detection was a highly improbable result. To summarize,
we take $F_{\nu,\rm lim} = 6$~mJy and $N_{\rm all-sky} = 34$.

\subsection{VLA Low-Galactic-Latitude Survey (5 GHz)}

\citet{Ofek+11} report a survey with a footprint of 2.66 deg$^2$ comprising 16
epochs. The spacing of the epochs (their Table 3) is such that only 2 of them
are independent on 100-day time scales, so we adopt A = 5.32 deg$^2$,
significantly smaller than the naive effective area of 41 deg$^2$. An
8$\sigma$ source was reported that was present in the very first epoch and
became undetectable 3 days later. Since this evolution appears unrealistically
rapid, this source was probably spurious even though it passed various quality
checks, a conclusion shared by other groups \citep{Frail+12}. \citet{Ofek+11}
report a representative flux density limit of 1.8 mJy (their section 9.1), so
we take $F_{\nu,\rm lim} = 1.8$~mJy and $N_{\rm all-sky} = 23270$. We note
that although we present the limit from this survey with our 3~GHz results,
the survey was conducted at 5~GHz.

\subsection{VLA Calibrator Field Survey (5 \& 8 GHz)}

\citet{Bower+07} report the results of a 944-epoch survey for
transient sources with archival data from the VLA spanning 22 years
with a typical epoch separation of 7 days. We update the results to
account for the improved re-analysis of \citet{Frail+12}. At a 100 day
cadence, the 22 years of observations represent only 80 independent
epochs, so the effective area is reduced by a factor of $\sim$ 10
compared to a naive estimate. Observations were performed at 4.9 and
8.4 GHz, and again we show this survey with our 3~GHz results.

\citet{Frail+12} conclude that there is only one transient in the data set,
with a flux density of 0.6 mJy at 4.9 GHz and a significance of 5.8$\sigma$.
It only appears in one epoch. As with the \citet{Jaeger+12} transient, the
detected event is too fast and implies a rate far above our predictions, but
we treat it as real. \citet{Bower+07} report effective areas as a function of
the flux density limit and the characteristic time scale (their Tables 5--7).
We employ the limit closest to the detected transient flux density of 0.56
mJy. On a 2-month time scale, their Table 6 gives an effective area of 4.97
deg$^2$ with 120 epochs. Scaling by 80/120 (i.e., converting $\Omega_s$ to use
the definition of \autoref{e.omegalarge} rather than \autoref{e.omegasmall})
gives a corrected effective area of 3.3 deg$^2$. We thus take $F_{\nu,\rm lim}
= 0.56$~mJy and $N_{\rm all-sky} = 12500$.

\bibliographystyle{yahapj}
\bibliography{ms}

\end{document}